\documentclass[aps,pra,noshowpacs,superscriptaddress,twocolumn]{revtex4-1}

\usepackage{mathrsfs}
\usepackage{amsfonts}
\usepackage{amssymb}
\usepackage{amsmath}
\usepackage{bm}
\usepackage{graphicx}
\usepackage{sidecap}
\usepackage{color}
\usepackage[colorlinks=true,citecolor=blue,linkcolor=magenta]{hyperref}
\usepackage{subeqnarray}
\usepackage{verbatim}
\usepackage{ulem}
\usepackage{diagbox}

\newcommand{\ket}[1]{\vert #1 \rangle}
\newcommand{\bra}[1]{\langle #1 \vert}
\newcommand{\ketbra}[2]{\vert #1 \rangle \langle #2 \vert}
\newcommand{\braket}[2]{\langle #1 \vert #2 \rangle}

\begin{document}
\title{Holonomic surface codes for fault-tolerant quantum computation}
\author{Jiang Zhang}
\affiliation{Quantum Physics and Quantum Information Division,
Beijing Computational Science Research Center, Beijing 100193, China}
\author{Simon J. Devitt}
\affiliation{RIKEN Center for Emergent Matter Science (CEMS), 2-1 Hirosawa, Wako, Saitama 351-0198, Japan}
\affiliation{Center for Engineered Quantum Systems, Department of Physics and Astronomy, Macquarie University, Sydney, New South Wales 2109, Australia}
\affiliation{Turing Inc., 2601 Dana Street, Berkeley, California 94704, USA}
\author{J. Q. You}
\thanks{jqyou@csrc.ac.cn}
\affiliation{Quantum Physics and Quantum Information Division,
Beijing Computational Science Research Center, Beijing 100193, China}
\author{Franco Nori}
\affiliation{RIKEN Center for Emergent Matter Science (CEMS), 2-1 Hirosawa, Wako, Saitama 351-0198, Japan}
\affiliation{Physics Department, The University of Michigan, Ann Arbor, Michigan 48109-1040, USA}

\begin{abstract}
Surface codes can protect quantum information stored in qubits from local errors as long as the per-operation error rate
is below a certain threshold. Here we propose holonomic surface codes by harnessing the quantum holonomy of the system.
In our scheme, the holonomic gates are built via auxiliary qubits rather than the auxiliary levels in multilevel
systems used in conventional holonomic quantum computation.
The key advantage of our approach is that the auxiliary qubits are in their ground state before and after each gate
operation, so they are not involved in the operation cycles of surface codes. This provides an advantageous
way to implement surface codes for fault-tolerant quantum computation.
\end{abstract}
\maketitle

\section{Introduction}
Quantum error correction (QEC) can protect quantum information from detrimental noise by encoding a logical qubit
with a set of physical qubits~\cite{shor,steane96,laflamme96}. Recently, remarkable progress has been achieved in
the experimental demonstration of QEC on small-size codes~\cite{yao12,barends14,nigg14,cor15,kelly15}.
A prototypical approach to QEC is the surface code~\cite{terhal15,bravyi98,dennis02,fowler12pra},
defined on a two-dimensional (2D) qubit lattice~\cite{sc97}. One attractive feature of these surface codes is
that only nearest-neighbor interactions are involved, which facilitates their experimental construction.
Another appealing feature is their appreciable tolerance to local errors~\cite{wang03,rau06,rau07prl,rau07njp,fowler12prl}.
According to a recent estimation of this tolerance with a balanced-noise model, the fidelities of every single- and
two-qubit gate for surface codes should be larger than $99.4\%$~\cite{stephens14}, so as to satisfy the fault-tolerant
threshold.

Current quantum control technology has made it experimentally accessible to reach this surface-code threshold for
fault tolerance~\cite{barends14}. However, quantum gates with higher fidelities are still needed since this could
allow one to greatly reduce the many number of physical qubits needed for encoding a logical qubit. For dynamic quantum
gates, stochastic Pauli errors are important~\cite{stephens14}, but they are less important for holonomic quantum
gates because the holonomic gates are robust against small stochastic fluctuations in the Hamiltonian of the
system (see the Appendices). Now, imperfect control of the Hamiltonian during gate operations may become the main
source of errors in the quantum-holonomy approach. Here, we propose holonomic surface codes, where the errors
caused by imperfect control can also be suppressed.

Non-Abelian geometric phases (i.e., quantum holonomy)~\cite{paolo1999,duan01,sjoqvist12,xu12,mousolou,gu15,feng13,zu14}
can be used to build quantum gates with higher fidelities, but existing holonomic schemes are based on multilevel
(usually three- or four-level) quantum systems~\cite{sjoqvist12,xu12,mousolou,gu15}. A direct application of these
schemes to surface codes is problematic because the required projective measurement on a multilevel quantum system
can have a state collapse to the auxiliary level, in addition to the levels used for a qubit. To overcome this problem,
here the holonomic gates are built with the help of auxiliary qubits rather than auxiliary levels used in the original
non-Abelian methods.
The required multilevel structure is provided by the Hilbert space spanned by both the target and auxiliary qubits together.
The key advantage of our method is that the auxiliary qubits are in their {\it ground states before and after}
each gate operation, so they are not involved in the operation cycles of surface codes.
Moreover, the construction of a holonomic gate only needs a Hamiltonian with nearest-neighbor $XY$-type interactions.
This is another advantage due to its accessibility in real systems.
Thus, our approach provides an advantageous way to implement surface codes for fault-tolerant quantum computation.

\section{Dynamic and holonomic surface codes}
\begin{figure}[t]
\centering
\includegraphics[width=0.48\textwidth]{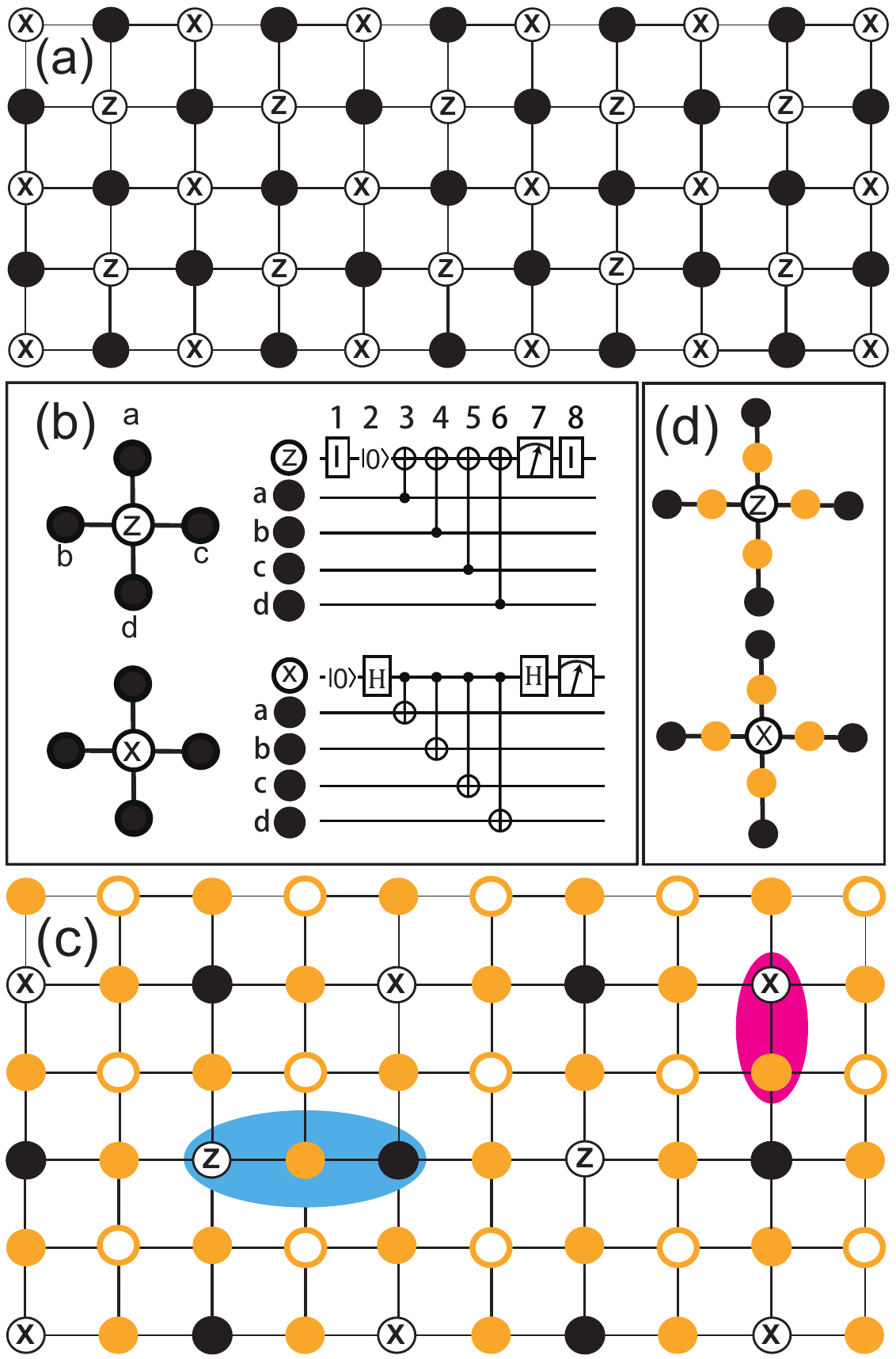}
\caption{(a) A 2D lattice implementation of dynamic surface codes. (b) The stabilizer units (left) and
syndrome extraction circuits (right) for measurement qubits $Z$ and $X$ in (a).
(c) A 2D lattice implementation of holonomic surface codes, where we introduce two types of auxiliary qubits,
shown as orange solid and open circles, respectively. (d) The stabilizer units for measurement qubits $Z$ and $X$ in (c).}
\label{lattice}
\end{figure}
Surface codes were studied on a square lattice of qubits \cite{fowler12pra}, with each qubit acting as either
a ``measurement" qubit (open circle) or a ``data" qubit (solid circle) [see Fig.~\ref{lattice}(a)].
The measurement qubit, denoted by $Z$ or $X$, has a stabilizer $Z_{abcd}\equiv\sigma_z^a\sigma_z^b\sigma_z^c\sigma_z^d$
or $X_{abcd}\equiv\sigma_x^a\sigma_x^b\sigma_x^c\sigma_x^d$, where $\sigma_z^i$ ($\sigma_x^i$), with $i=a$,$b$,$c$,$d$,
are Pauli operators acting on the four data qubits adjoining the measurement qubit [Fig.~\ref{lattice}(b)].
At the boundary of the lattice, the stabilizer is reduced to having three Pauli operators acting on the three
data qubits adjoining the measurement qubit. These stabilizers commute with each other because two (zero)
data qubits are shared by two neighboring (non-neighboring) stabilizers.
With projective measurements on all the measurement qubits, the state $\ket{\psi}$ of all the data qubits satisfies
$Z_{abcd}|\psi\rangle=z_{abcd}|\psi\rangle$ and $X_{abcd}|\psi\rangle=x_{abcd}|\psi\rangle$
for all stabilizers, with $z_{abcd}$, $x_{abcd}=\pm 1$.

A $Z$($X$)stabilizer requires a Hamiltonian with four-body interactions. This is impossible because four-body interactions
do not occur in natural systems. Thus, a sequence of single- and two-qubit operations, plus a projective measurement,
are utilized to achieve the stabilizer combined with the required projective measurement [see Fig.~\ref{lattice}(b)].
Then, as shown in \cite{fowler12pra}, logical qubits can be created by making holes (i.e., defects) inside the lattice,
and the operations on logical qubits for quantum computing can be decomposed into single- and two-qubit operations on
physical qubits. However, in order to implement large-scale quantum computing with surface codes, the fidelity threshold
of any operation on physical qubits should be above $99.4\%$~\cite{stephens14}. For a large system with many qubits,
this is extremely difficult to achieve when dynamic operations are employed.

Besides dynamic operations, one can harness the geometric nature of the quantum system to build quantum operations
(on physical qubits) with extremely high fidelities.
The implementation of surface codes involves noncommutative operations on physical qubits, so the employed geometric
quantum gates should be constructed with quantum holonomy.
In the existing holonomic approach, multilevel quantum systems are utilized to implement the holonomic operations
on qubits. However, it is problematic to directly apply this approach to surface codes since the required projective
measurement in surface codes can collapse the state of the system to the auxiliary levels, in addition to the qubit levels.

Here we propose a square lattice of qubits [Fig.~\ref{lattice}(c)] where, besides data and measurement qubits,
some additional qubits (shown as orange circles) are used as auxiliary qubits. For each pair of adjacent data and
measurement qubits, an auxiliary qubit (orange solid circle) lies in between, so each stabilizer unit contains four
auxiliary qubits [Fig.~\ref{lattice}(d)].
We assume that the Pauli-$x$ and Pauli-$y$ Hamiltonian of each single physical qubit are available. In addition,
an $XY$-type interaction Hamiltonian between nearest-neighbor physical qubits is required, which reads
\begin{equation}\label{E1}
\hat{H}_{jk}^{XY}=\frac{J_{jk}}{2}(\sigma^j_x\sigma^k_x+\sigma^j_y\sigma^k_y),
\end{equation}
where $J_{jk}$ is the coupling strength and $\sigma^j_x$ ($\sigma^j_y$) represents the Pauli-$x$ (-$y$) operator
on the $j$th physical qubit.
Also, we assume that the Hamiltonian for each single qubit can be tuned to zero. Moreover, we can turn on the
coupling strengths  $J_{jk}$ for a certain time, so as to satisfy the cyclic condition for achieving holonomic gates.
With this lattice, single- and two-qubit holonomic operations can be implemented using physical
qubits (see Sec.~III below), instead of employing multilevel systems. This makes it possible to realize holonomic
surface codes.

\section{Holonomic operations for surface codes}
We first explain how quantum holonomy can arise in nonadiabatic unitary evolution. Consider a quantum system with
Hamiltonian $\hat{H}$, which spans a Hilbert space containing a subspace $\mathcal{L}$ spanned by a set of
basis states $\{\ket{\psi_k}\}_{k=1}^m$, where $m$ is the dimension of $\mathcal{L}$.
We can define a projection operator $P_{\mathcal{L}}=\sum_{k=1}^{m}\ketbra{\psi_k}{\psi_k}$ for $\mathcal{L}$.
Then, nonadiabatic quantum holonomy acting on $\mathcal{L}$ can be realized if two conditions are
satisfied \cite{sjoqvist12,xu12}:

(i) $\hat{H}$ vanishes in the evolving subspace $P_{\mathcal{L}}(t)=U(t)P_{\mathcal{L}}U^\dagger(t)$.

(ii) The subspace $\mathcal{L}$ evolves in a cyclic manner following
$\mathcal{L}(t)\equiv\mathrm{Span}\{U(t)\ket{\psi_k}\}_{k=1}^m$,
where $U(t)=\mathrm{T}\exp[-i\int_{0}^{t}\hat{H}(t')\mathrm{d}t']$, with $\mathrm{T}$ denoting time ordering.

\begin{figure}[t]
\centering
\includegraphics[width=0.4\textwidth]{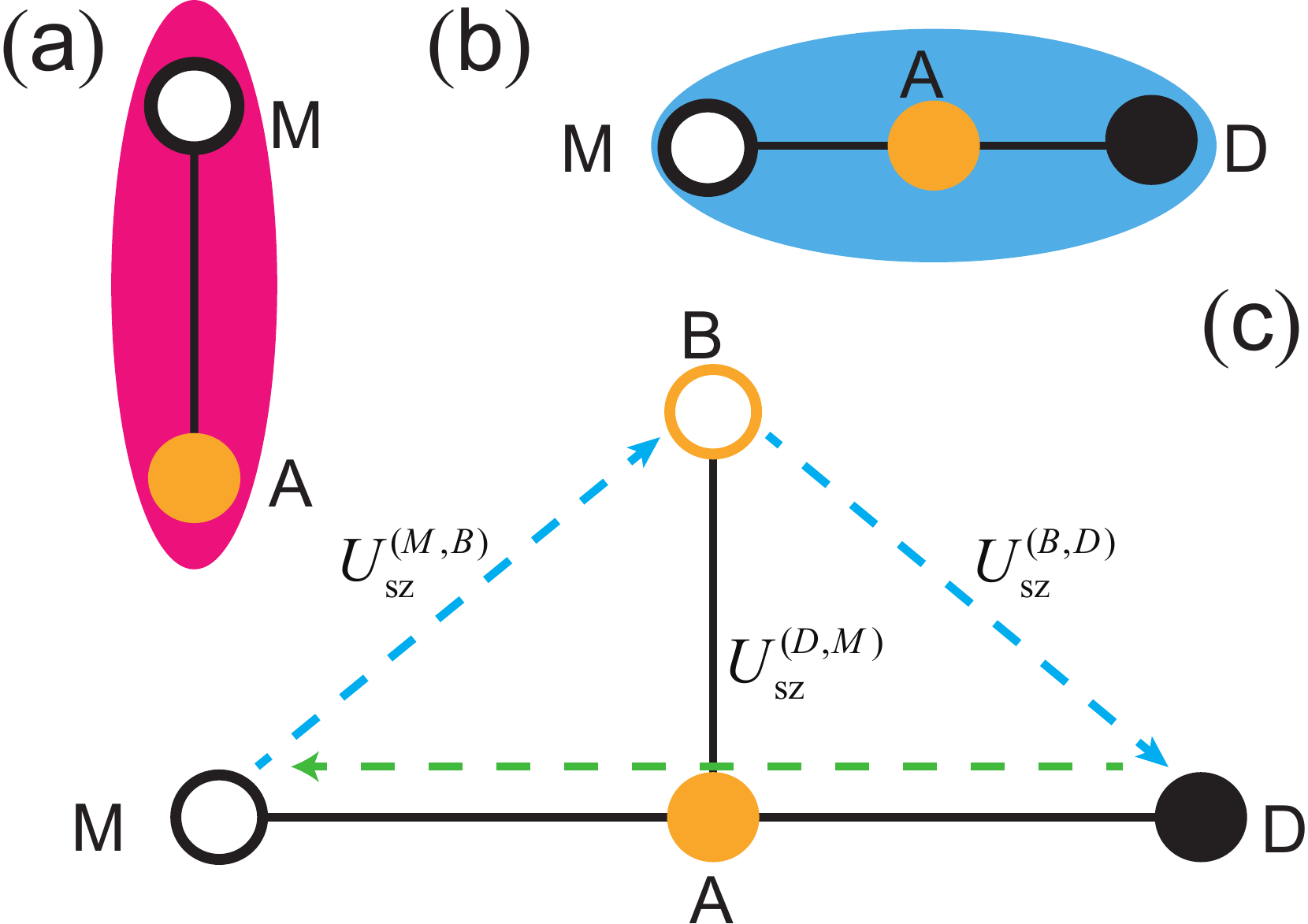}
\caption{(a)~Implementation of holonomic single-qubit operations.
(b)~Implementation of the holonomic quantum gate $U_{\mathrm sz}$ between qubits $M$ and $D$.
(c)~Implementation of the holonomic SWAP gate between qubits $M$ and $D$ via three holonomic
two-qubit $U_{\mathrm sz}$ gates (first between qubits $M$ and $B$, then between qubits $D$ and $M$,
and finally between qubits $B$ and $D$).}
\label{qubitgates}
\end{figure}

To construct a holonomic single-qubit gate on a target (either data or measurement) qubit, we need a nearby auxiliary
qubit interacting with it via an $XY$-type interaction [see Fig.~\ref{qubitgates}(a)]. Specifically,
we use ``1" (``2") to denote the auxiliary (target) qubit. The Hamiltonian of the two qubits is
\begin{equation}\label{h1}
  \hat{H}_1=\frac{J_1}{2}(\cos\beta\sigma_x^1+\sin\beta\sigma_y^1)
+\frac{J_{12}}{2}(\sigma_x^1\sigma_x^2+\sigma_y^1\sigma_y^2),
\end{equation}
where $\sigma_x^1$ ($\sigma_y^1$) is a Pauli matrix acting on qubit 1 and $\beta$ is a constant.
By setting the parameters $J_1=J_0(t)\sin\theta$ and $J_{12}=J_0(t)\cos\theta$,
the time-evolution operator $U_1(t)$ generated by $\hat{H}_1(t)$ can be obtained as (see Appendix A)
\begin{equation}\label{U1}
  U_1(t)=\left(
           \begin{array}{cc}
             V_0\cos(a_tD)V_0^\dagger & -iV_0\sin(a_tD)V_1^\dagger \\
             -iV_1\sin(a_tD)V_0^\dagger & V_1\cos(a_tD)V_1^\dagger \\
           \end{array}
         \right),
\end{equation}
where $a_t\!=\!\int_{0}^{t}J_0(s)\mathrm{d}s$, and
\begin{eqnarray}\label{E4}
V_0&=&\left(
             \begin{array}{cc}
               \sin\frac{\theta}{2} & e^{-i\beta}\cos\frac{\theta}{2} \\
               e^{i\beta}\cos\frac{\theta}{2} & -\sin\frac{\theta}{2} \\
             \end{array}
           \right), \nonumber\\
D&=&\left(
            \begin{array}{cc}
                      \cos^2\frac{\theta}{2} & 0 \\
                      0 & \sin^2\frac{\theta}{2} \\
                    \end{array}
                  \right), \\
V_1^\dagger&=&\left(
          \begin{array}{cc}
            \cos\frac{\theta}{2} & e^{-i\beta}\sin\frac{\theta}{2} \\
            e^{i\beta}\sin\frac{\theta}{2} & -\cos\frac{\theta}{2} \\
          \end{array}\nonumber
        \right).
\end{eqnarray}

We divide the 4D Hilbert space of $\hat{H}_1$ into two subspaces spanned by
$\mathcal{L}_0=\mathrm{Span}\{\ket{00},\ket{01}\}$ and $\mathcal{L}_1=\mathrm{Span}\{\ket{10},\ket{11}\}$,
and accordingly define two projection operators
$P_0=\ketbra{00}{00}+\ketbra{01}{01}$ and $P_1=\ketbra{10}{10}+\ketbra{11}{11}$.
By choosing a certain time $t=\tau$ such that $\sin(a_\tau D)=\mathrm{diag}\{0,0\}$
but $\cos(a_\tau D)=\mathrm{diag}\{-1,1\}$ (this condition can always be satisfied as long as
$\theta\neq\frac{1+2n}{2}\pi$), we obtain the final evolution operator,
\begin{equation}
U_1(\mathcal{L}_0,\mathcal{L}_1)=\ketbra{0}{0}\otimes V(\mathcal{L}_0)+\ketbra{1}{1}\otimes V(\mathcal{L}_1),
\end{equation}
where
\begin{eqnarray}
V(\mathcal{L}_k)&=&-V_k\sigma_zV_k^\dagger \nonumber\\
&=&(-1)^{k}\cos\theta\sigma_z-\sin\theta(\cos\beta\sigma_x+\sin\beta\sigma_y)~~~~
\end{eqnarray}
is a single-qubit gate acting on qubit 2.
Given that qubit 1 is initialized in $\ket{0}$,  the gate on qubit 2 is a Hadamard gate $H$ if
$\theta=\pi/4$ and $\beta=\pi$.
A single-qubit rotation around the $x$axis, $R_x(\alpha)=\exp(-i\alpha\sigma_x)$, can be obtained by applying
$V(\mathcal{L}_0)$ twice, during which $\theta=\pi/4$ and $\beta=\pi/2$ for the first cycle,
while $\theta=\pi/4+\alpha$ and $\beta=\pi/2$ for the second one.
Another rotation around the $z$axis, $R_z(\alpha)=\exp(-i\alpha\sigma_z)$, is also available since
$HR_x(\alpha)H=R_z(\alpha)$. Note that condition (i) is satisfied because
\begin{equation}
P_k(t)\hat{H}_1P_k(t)=U_1(t)P_k\hat{H}_1P_kU_1^\dagger(t)=0,
\end{equation}
where $[\hat{H}_1,U_1(t)]=0$ is used. Also, because
\begin{equation}
\mathcal{L}_j(\tau)=U_1(\tau)\mathcal{L}_j(0)=\mathcal{L}_j(0)\quad (j=0,1),
\end{equation}
condition (ii) is satisfied. Thus, $V(\mathcal{L}_k)$ is a holonomic single-qubit gate.

To construct a holonomic two-qubit gate on a pair of measurement ($M$) and data ($D$) qubits,
we introduce an auxiliary ($A$) qubit lying in between and interacting with both of them [see Fig.~\ref{qubitgates}(b)].
Hereafter, qubits $M$, $D$, and $A$ are also denoted as qubit 1, 2, and 3, respectively.
In this case, we turn off the single-qubit Hamiltonian and turn on the interactions.
Then the corresponding Hamiltonian is
\begin{equation}
\hat{H}_2=\hat{H}_{13}^{XY}+\hat{H}_{23}^{XY},
\end{equation}
which has four invariant subspaces:
\begin{eqnarray}
\mathcal{S}_1&\equiv&\mathrm{Span}\{\ket{000}\},~~
\mathcal{S}_2\equiv\mathrm{Span}\{\ket{001},\ket{010},\ket{100}\}, \nonumber\\
\mathcal{S}_3&\equiv&\mathrm{Span}\{\ket{110},\ket{101},\ket{011}\},~~
\mathcal{S}_4\equiv\mathrm{Span}\{\ket{111}\}.\nonumber
\end{eqnarray}
Here and hereafter, for the state $\ket{ijk}$, we use $i$ for qubit 1, $j$ for qubit 2, and $k$ for qubit 3.
In the subspaces $\mathcal{S}_1$ and $\mathcal{S}_4$, $\hat{H}_2$ simply reduces to
$\hat{H}_{\mathcal{S}_1}=\hat{H}_{\mathcal{S}_4}=0$, but it reduces to
\begin{equation}\label{eq3}
  \hat{H}_{\mathcal{S}_2}=\hat{H}_{\mathcal{S}_3}=\left(
      \begin{array}{ccc}
        0 & J_{23} & J_{13} \\
        J_{23} & 0 & 0 \\
        J_{13} & 0  & 0 \\
      \end{array}
    \right)
\end{equation}
in the subspaces $\mathcal{S}_2$ and $\mathcal{S}_3$.

We first focus on the evolution in $\mathcal{S}_2$ owing to $\hat{H}_{\mathcal{S}_2}$.
We use the basis states $\ket{010}$ and $\ket{100}$ as the logical $\ket{0}$ and $\ket{1}$ to define the qubit subspace
$\mathcal{S}_{2,q}(0)$, and $\ket{001}$ to define the auxiliary subspace $\mathcal{S}_{2,a}(0)$.
Here we choose $J_{23}$ and $J_{13}$ as the controllable parameters and set $J_{23}=\Omega(t)\sin(\theta/2)$ and
$J_{13}=\Omega(t)\cos(\theta/2)$,
where $\Omega(t)$ describes the envelope, and $\theta$ is a time-independent parameter.
By turning on $\hat{H}_{\mathcal{S}_2}$, the qubit subspace $\mathcal{S}_{2,q}(0)$ evolves to $\mathcal{S}_{2,q}(t)$
spanned by the ordered basis $\{U_{\mathcal{S}_2}(t)\ket{010},U_{\mathcal{S}_2}(t)\ket{100}\}$,
where $U_{\mathcal{S}_2}(t)=\exp[-i\int_0^t\hat{H}_{\mathcal{S}_2}(t^\prime)\mathrm{d}t^\prime]$.
Meanwhile, the auxiliary subspace $\mathcal{S}_{2,a}(0)$ evolves to $\mathcal{S}_{2,a}(t)$ spanned by
$U_{\mathcal{S}_2}(t)\ket{001}$. With $\mathcal{S}_{2,q}$ and $\mathcal{S}_{2,a}$ evolved for a certain time $\tau$,
such that $\int_0^\tau\Omega(t^\prime)\mathrm{d}t^\prime=\pi$, the resulting unitary operator becomes
\begin{equation}\label{us2}
  U_{\mathcal{S}_2}(\tau)=\left(
                      \begin{array}{ccc}
                        -1 & 0 & 0 \\
                        0 & \cos\theta & -\sin\theta \\
                        0 & -\sin\theta & -\cos\theta \\
                      \end{array}
                    \right).
\end{equation}
Thus, we obtain a negative identity operator
$U_{\mathcal{S}_{2,a}}(\tau)=-I_a$ on the subspace $\mathcal{S}_{2,a}(0)$, and a quantum gate
\begin{equation}
U_{\mathcal{S}_{2,q}}(\tau)=\cos\theta\sigma_z-\sin\theta\sigma_x
\end{equation}
on the subspace
$\mathcal{S}_{2,q}(0)$.
Note that condition (i) is satisfied, since $U_{\mathcal{S}_2}(t)$ commutes with $\hat{H}_{\mathcal{S}_2}$, so
\begin{eqnarray}
P_{\mathcal{S}_{2,k}}(t)\hat{H}_{\mathcal{S}_2}P_{\mathcal{S}_{2,k}}(t)&=&
U_{\mathcal{S}_2}(t)P_{\mathcal{S}_{2,k}}\hat{H}_{\mathcal{S}_2}P_{\mathcal{S}_{2,k}}U_{\mathcal{S}_2}^\dagger(t) \nonumber\\
&=&0 \quad (k=a,q).
\end{eqnarray}
Condition (ii) is also satisfied, because
\begin{align}
\mathcal{S}_{2,a}(\tau)&\equiv\mathrm{Span}\{U_{\mathcal{S}_{2,a}}(\tau)\ket{001}\} \nonumber\\
&=\mathrm{Span}\{\ket{001}\}, \nonumber\\
\mathcal{S}_{2,q}(\tau)&\equiv\mathrm{Span}\{U_{\mathcal{S}_{2,q}}(\tau)\ket{001},U_{\mathcal{S}_{2,q}}(\tau)\ket{100}\}
\nonumber \nonumber\\
&=\mathrm{Span}\{\ket{001},\ket{100}\}.
\end{align}
Therefore, the gate $U_{\mathcal{S}_2}(\tau)$ is holonomic.
Similarly, evolving $\mathcal{S}_3$ for the same time interval $\tau$, we obtain a holonomic gate
$U_{\mathcal{S}_{3,a}}(\tau)=-I_a$ on $\mathcal{S}_{3,a}\equiv\mathrm{Span}\{\ket{110}\}$, and another holonomic
gate on $\mathcal{S}_{3,q}\equiv\mathrm{Span}\{\ket{101},\ket{011}\}$ which reads
\begin{equation}
U_{\mathcal{S}_{3,q}}(\tau)=\cos\theta\sigma_z-\sin\theta\sigma_x.
\end{equation}

The state of qubits $M$ and $D$ is generally an arbitrary two-qubit pure state, $\sum_{m,n=0,1} a_{mn}\ket{mn}$.
We initialize qubit $A$ in $\ket{0}$. Then, the initial state of the three qubits  reads
\begin{equation}
\ket{\psi}_i=(a_{00}\ket{00}+a_{01}\ket{01}+a_{10}\ket{10}+a_{11}\ket{11})\otimes\ket{0}.
\end{equation}
When the cyclic condition is satisfied, the state of qubit $A$ returns to $\ket{0}$ and the final state of the
three qubits becomes
\begin{equation}
\ket{\psi}_f=[a_{00}\ket{00}+U_{\mathcal{S}_{2,q}}(a_{01}\ket{01}+a_{10}\ket{10})-a_{11}\ket{11}]\otimes\ket{0}.
\end{equation}
The corresponding holonomic two-qubit gate on the physical qubits $M$ and $D$ is given by
\begin{equation}
U_2=I_{00}\oplus U_{\mathcal{S}_{2,q}}\oplus (-I_{11}),
\end{equation}
where $I_{00(11)}$ is an identity operator acting on $\ket{00}$ ($\ket{11}$).
Choosing $\theta=3\pi/2$, we obtain a nontrivial holonomic two-qubit gate,
\begin{equation}\label{u2}
  U_{\mathrm{sz}}^{(M,D)}=\left(
        \begin{array}{cccc}
          1 & 0 & 0 & 0 \\
          0 & 0 & 1 & 0 \\
          0 & 1 & 0 & 0 \\
          0 & 0 & 0 & -1 \\
        \end{array}
      \right).
\end{equation}
Note that $U_{\mathrm{sz}}$ transforms $\ket{0}\otimes(a\ket{0}+b\ket{1})$ to $(a\ket{0}+b\ket{1})\otimes\ket{0}$,
and vice versa. A SWAP gate between qubits $M$ and $D$ can be realized in three steps by employing an extra auxiliary
qubit ($B$) initialized in $\ket{0}$, in addition to the one ($A$) between qubits
$M$ and $D$ [see Fig.~\ref{qubitgates}(c)]:
\begin{equation}
U_{\mathrm{SWAP}}^{(M,D)}=U_{\mathrm{sz}}^{(B,D)}U_{\mathrm{sz}}^{(D,M)}U_{\mathrm{sz}}^{(M,B)}.
\end{equation}
A control-Z (CZ) gate is obtained as
\begin{equation}
U_{\mathrm{CZ}}^{(M,D)}=U_{\mathrm{sz}}^{(M,D)}U_{\mathrm{SWAP}}^{(M,D)}.
\end{equation}
Accordingly, a control-NOT (CNOT) gate is given by
\begin{equation}
U_{\mathrm{CNOT}}^{(M,D)}=HU_{\mathrm{CZ}}^{(M,D)}H,
\end{equation}
where $H$ is the Hadamard gate on qubit $D$.

So far, we have shown how to construct all the holonomic single- and two-qubit gates needed for surface codes.
In our lattice [Fig.~\ref{lattice}(c)], there is always an auxiliary qubit below every target qubit,
so we can implement holonomic single-qubit gates on all target qubits at the same time. Also, the two-qubit gate
architecture in Fig.~\ref{qubitgates}(c) exists for every pair of measurement and data qubits in our lattice.
For each measurement $X$ (or $Z$) qubit, we can implement a holonomic CNOT gate on this qubit and its nearest-neighbor
data qubit $i$, where $i=a$,$b$,$c$,$d$ [cf.~Fig.~\ref{qubitgates}(b)]. Among these eight types of CNOT gates,
we can simultaneously implement all CNOT gates of the same type. Thus, all the $X$ (or $Z$) syndrome extraction circuits
can be implemented in parallel.

\section{Gate robustness to errors and auxiliary-qubit resetting}
In Appendix B, we show how Pauli errors spread in holonomic quantum gates. In fact, as a distinct advantage,
holonomic quantum gates are robust against small stochastic fluctuations (see Appendix C), including also stochastic
Pauli errors. Below we consider the area error due to  imperfect control of the Hamiltonian because now it may become
the main source of errors. When an area error $\delta$ is included, the cyclic condition
$\int_0^\tau\Omega(s)\mathrm{d}s=\pi$ for a holonomic $U_{\mathrm{sz}}$ gate becomes
$\int_0^\tau\Omega(s)\mathrm{d}s=\pi+\delta$.
The final state of the employed three qubits is
\begin{equation}
\ket{\psi^{(r)}}_f=\ket{\psi}_0\otimes\ket{0}+i\sin\delta\ket{\psi}_1\otimes\ket{1},
\end{equation}
where
\begin{eqnarray}
\ket{\psi}_0&=&a_{00}\ket{00}+(a_{01}\sin^2\frac{\delta}{2}+a_{10}\cos^2\frac{\delta}{2})\ket{01} \nonumber\\
&&+(a_{10}\sin^2\frac{\delta}{2}+a_{01}\cos^2\frac{\delta}{2})\ket{10}-a_{11}\cos\delta\ket{11}, \nonumber\\
\ket{\psi}_1&=&\frac{\sqrt{2}}{2}(a_{01}+a_{10})\ket{00}+\frac{\sqrt{2}}{2}a_{11}\ket{10} \nonumber\\
&&+\frac{\sqrt{2}}{2}a_{11}\ket{01}.
\end{eqnarray}
The lower bound of the fidelity for the holonomic $U_{\mathrm{sz}}$ gate can be written as (see Appendix D)
\begin{equation}
F_h=1-\frac{\delta^4}{1-\delta^2+\delta^4}.
\end{equation}
For the holonomic single-qubit gate (e.g., Hadamard gate), its fidelity is even higher (see also Appendix D).

Owing to the error $\delta$, the two target qubits become entangled with the auxiliary qubit in the real final state
$\ket{\psi^{(r)}}_f$. To reset the auxiliary qubit to its ground state $\ket{0}$, one needs to implement a measurement
with $\sigma_z$ on the auxiliary qubit.
When $\ket{0}$ is obtained for the auxiliary qubit, the state of the two target qubits is collapsed to $\ket{\psi}_0$
with a very high probability. Although $\ket{\psi}_0\otimes\ket{0}$ deviates a bit from the ideal final state
$\ket{\psi^{(i)}}_f\equiv\ket{\psi^{(r)}}_f|_{\delta=0}$, it is perfectly correctable by surface codes as long as
$\delta$ is small enough. However, when $\ket{1}$ is obtained for the auxiliary qubit, the state of the two target
qubits is collapsed to $\ket{\psi}_1$ with a very low probability. Here, $\ket{\psi}_1\otimes\ket{1}$ is free of
the area error $\delta$ and can be converted to $\ket{\psi^{(i)}}_f$ via holonomic operations (see Appendix E).

Also, we can use $XY$ interactions between the measurement and data qubits to directly build a dynamic iSWAP gate
by turning on
\begin{equation}
\hat{H}_d=\frac{\Omega(t)}{2}(\sigma_x^M\sigma_x^D+\sigma_y^M\sigma_y^D)
\end{equation}
for a time $t$, so that $\int_{0}^{t}\Omega(s)\mathrm{d}s=\frac{3\pi}{2}$. We calculate the fidelity of the
dynamic iSWAP gate by considering the area error $\int_{0}^{t}\Omega(s)\mathrm{d}s=\frac{3\pi}{2}+\delta$.
The corresponding lower bound of the gate fidelity is (see Appendix D)
\begin{equation}
F_d=1-\frac{\delta^2}{2}.
\end{equation}

\begin{figure}[t]
\centering
\includegraphics[width=0.48\textwidth]{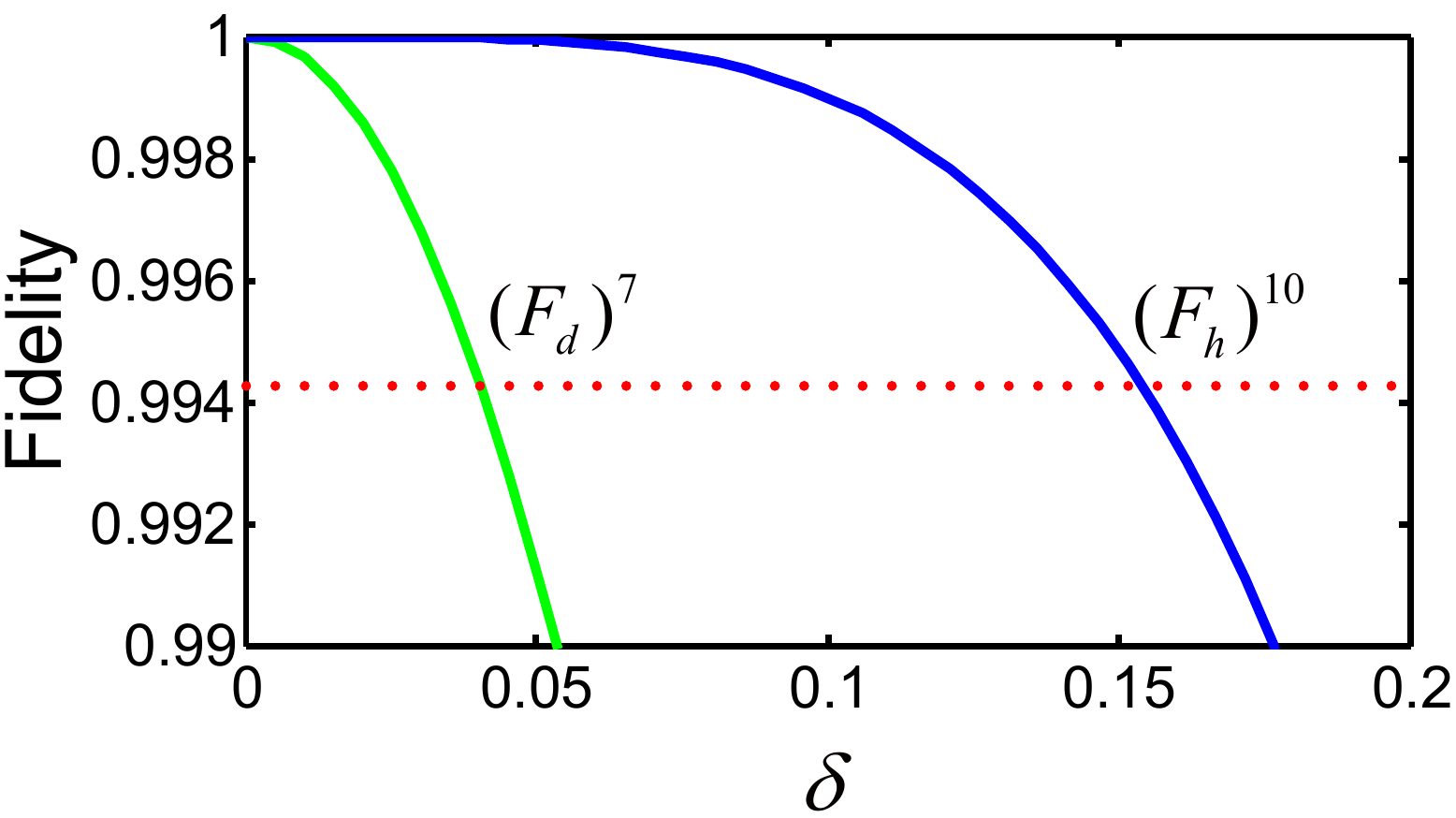}
\caption{The lower bound of the fidelity for the holonomic CNOT gate (blue curve) and that for the dynamic CNOT gate (green curve) vs the area error $\delta$. For $\delta\lesssim0.152$, the fidelity of the holonomic CNOT is larger than the threshold $99.4\%$ (marked by the red dotted line), but it requires $\delta\lesssim0.041$ for the dynamic one.}
\label{fidelity}
\end{figure}

A holonomic CNOT gate can be constructed in ten steps, including four $U_{\mathrm{sz}}$'s, four initializations,
and two Hadamards. With an iSWAP gate, a dynamic CNOT gate can be realized in seven steps~\cite{schuch,Tanamoto}.
According to the balanced-noise model, the lower bounds of the gate fidelities for holonomic and dynamic CNOT gates
are then given by $(F_h)^{10}$ and $(F_d)^7$, respectively.
Figure~\ref{fidelity} shows the numerical results for both of them. It is clear that the holonomic gate fidelity
decreases more slowly than the dynamic one when increasing $\delta$, indicating that the holonomic gate is more robust
than the dynamic gate.

\section{Discussion and conclusion}
We have proposed a method to implement surface codes by harnessing quantum holonomy,
where the holonomic gates on target qubits are built with the help of auxiliary qubits. After each gate operation,
the measurement on the auxiliary qubit can provide a heralding signal which can be used to improve gate fidelity.
In Ref.~\cite{mousolou}, the same Hamiltonian as in Eq.~(\ref{h1}) was used to achieve holonomic
single-qubit gates.
In Ref.~\cite{gu15}, the three-level systems, instead of two-level systems, were used to realize the holonomic
single-qubit gates, but only two levels of each three-level system were used to construct the holonomic two-qubit
gates and the Hamiltonian used is also the $XY$ type.
In our approach, all holonomic single- and two-qubit gates are achieved by employing auxiliary qubits instead of
using three-level systems. Moreover, we focus on achieving surface codes for fault-tolerant quantum computation
via the required holonomic single- and two-qubit gates, rather than the holonomic quantum computation considered
in, e.g., Refs.~\cite{mousolou} and  \cite{gu15}.

Compared with the original surface-code lattice, our holonomic scheme consumes more qubits to obtain the same
code distance. However, by paying such a price, our approach gains a significant advantage:~the holonomic two-qubit
gate is much more robust than the conventional dynamic two-qubit gate achieved with the same interaction Hamiltonian.
For example, it follows from Fig.~\ref{fidelity} that when $\delta=0.041$, the fidelity of the dynamic CNOT gate
just reaches the needed fidelity $99.4\%$ of the quantum gates for surface codes. However, for the same value of
$\delta$, the fidelity of the corresponding holonomic CNOT gate is about $99.9993\%$, which is much higher than $99.4\%$.
Moreover, even though this scheme requires an increase by a factor of 4 of the number of physical qubits,
we achieve a reduction of error caused by imperfect control by three orders of magnitude.
This is advantageous for constructing a logical qubit.
Given that the logical error rate of a surface-code qubit decreases roughly as $\approx(p/p_{\mathrm {th}})^{d}$,
where $d$ is the distance of the code and $p_{\mathrm {th}}\approx 0.6\%$ is the threshold, by reducing $p$ to
three orders of magnitude less than $p_{\mathrm {th}}$, the required distance of the code and hence the total number
of physical qubits required for a logical qubit will easily result in a net benefit.

\acknowledgments
{This work is supported by the National Key Research and Development Program of China (Grant No.~2016YFA0301200),
the National Basic Research Program of China (Grant No.~2014CB921401), the NSFC (Grant No.~11774022), and the NSAF (Grant No.~U1530401).
F.N. was partially supported by the RIKEN iTHES Project, MURI Center for Dynamic Magneto-Optics via the AFOSR Award
No.~FA9550-14-1-0040, the Japan Society for the Promotion of Science (KAKENHI), the IMPACT program of JST,
JSPS-RFBR Grant No.~17-52-50023, CREST Grant No.~JPMJCR1676, and the Sir John Templeton Foundation.
S.J.D. acknowledges support by the Australian Research Council Centre of Excellence in Engineered Quantum Systems
EQUS (Project~CE110001013).}

\appendix
\section{Derivation of the time evolution operator in Eq.~(\ref{U1})}

By setting the parameters $J_1=J_0(t)\sin\theta$ and $J_{12}=J_0(t)\cos\theta$, the Hamiltonian in Eq.~(\ref{h1})
becomes
\begin{equation}\label{ht}
  \hat{H}_1=J_0(t)\left(
              \begin{array}{cc}
                0 & T \\
                T^\dagger & 0 \\
              \end{array}
            \right)
\end{equation}
in the ordered basis $\{\ket{00},\ket{01},\ket{10},\ket{11}\}$, where
\begin{equation}
T=\left(
      \begin{array}{cc}
      \frac{1}{2}\sin\theta e^{-i\beta} & 0 \\
        \cos\theta & \frac{1}{2}\sin\theta e^{-i\beta} \\
      \end{array}
    \right)
\end{equation}
is a $2\times2$ time-independent matrix. Since $T$ is invertible, there is a unique singular value
decomposition~\cite{mousolou},
\begin{equation}\label{svd}
  T=V_0DV_1^\dagger,
\end{equation}
with
\begin{align}
&V_0=\left(
             \begin{array}{cc}
               \sin\frac{\theta}{2} & e^{-i\beta}\cos\frac{\theta}{2} \\
               e^{i\beta}\cos\frac{\theta}{2} & -\sin\frac{\theta}{2} \\
             \end{array}
           \right), \\
&D=\left(
                    \begin{array}{cc}
                      \cos^2\frac{\theta}{2} & 0 \\
                      0 & \sin^2\frac{\theta}{2} \\
                    \end{array}
                  \right),
\end{align}
and
\begin{equation}
V_1^\dagger=\left(
          \begin{array}{cc}
            \cos\frac{\theta}{2} & e^{-i\beta}\sin\frac{\theta}{2} \\
            e^{i\beta}\sin\frac{\theta}{2} & -\cos\frac{\theta}{2} \\
          \end{array}
        \right).
\end{equation}

According to Eq. (\ref{svd}), it is easy to check that
\begin{equation}
\hat{H}_1^n=J_0^n(t)\left(
\begin{array}{cc}
 0 & V_0D^nV_1^\dagger \\
   V_1D^nV_0^\dagger & 0 \\
    \end{array}
    \right)
\end{equation}
for odd $n$, and
\begin{equation}
\hat{H}_1^n=J_0^n(t)\left(
\begin{array}{cc}
V_0D^nV_0^\dagger & 0 \\
 0 & V_1D^nV_1^\dagger \\
  \end{array}
   \right)
\end{equation}
for even $n$. Then, the time-evolution operator $U_1(t)$ generated by $\hat{H}_1(t)$ can be obtained as
\begin{align}\label{ut}
  U_1(t) =& T\exp\{-i\int_{0}^{t}\hat{H}_1(t')\mathrm{d}t'\} \nonumber \\
  =&\sum_{n=\mathrm{even}}\frac{(-i\int_0^t \hat{H}_1(t')\mathrm{d}t')^n}{n!}\nonumber\\
  &+\sum_{n=\mathrm{odd}}\frac{(-i\int_0^t \hat{H}_1(t')\mathrm{d}t')^n}{n!} \nonumber \\
  =&\left(
           \begin{array}{cc}
             V_0\cos(a_tD)V_0^\dagger & -iV_0\sin(a_tD)V_1^\dagger \\
             -iV_1\sin(a_tD)V_0^\dagger & V_1\cos(a_tD)V_1^\dagger \\
           \end{array}
         \right),
\end{align}
where $T$ denotes the time ordering and $a_t\!=\!\int_{0}^{t}J_0(t')\mathrm{d}t'$.
This is Eq.~(3) in the main text.

\section{Error spreading due to stochastic Pauli errors}

In the proposed holonomic surface codes, we introduce auxiliary qubits to construct the required quantum gates,
so more physical qubits than in the dynamic surface codes are consumed for a given code distance.
Here we show how stochastic Pauli errors spread in holonomic quantum gates.

\subsection{Holonomic single-qubit gate}

As shown in the main text, a holonomic single-qubit gate on a target qubit is achieved with the help of an auxiliary
qubit. Thus, it involves two physical qubits. Also, one needs to prepare the auxiliary qubit in $\ket{0}$ before
implementing the gate operation and then perform a measurement on it after the gate operation.
The evolution operator for these two physical qubits reads
\begin{equation}\label{}
  U_1(\mathcal{L}_0,\mathcal{L}_1)=\ketbra{0}{0}\otimes V(\mathcal{L}_0)+\ketbra{1}{1}\otimes V(\mathcal{L}_1).
\end{equation}
It is evident that if the auxiliary qubit is prepared in the orthogonal state $\ket{1}$, a similar holonomic
single-qubit gate $V(\mathcal{L}_1)$ on the target qubit is achieved. We assume that a stochastic two-qubit
Pauli error occurs, which affects the two physical qubits and can be chosen from
$\{I\sigma_x,I\sigma_y,I\sigma_z,\sigma_xI,\sigma_x\sigma_x,\sigma_x\sigma_y,\sigma_x\sigma_z\cdots,\sigma_z\sigma_z\}$
(15 errors in total)~\cite{stephens14}.
When performing a measurement on the auxiliary qubit, if an incorrect measurement result is reported, only the
target qubit is affected and the error is not spread to other physical qubits.
Therefore, local Pauli errors do not spread out through a holonomic single-qubit gate.

\subsection{Holonomic two-qubit gate}

Also, as shown in the main text, a holonomic two-qubit gate on two target qubits is achieved with the help of
an auxiliary qubit. In the proposed holonomic surface codes, the basic holonomic two-qubit gate for constructing
the needed CNOT gate is
\begin{equation}\label{}
  U_{\mathrm{sz}}=\left(
        \begin{array}{cccc}
          1 & 0 & 0 & 0 \\
          0 & 0 & 1 & 0 \\
          0 & 1 & 0 & 0 \\
          0 & 0 & 0 & -1 \\
        \end{array}
      \right).
\end{equation}
Unlike $U_1$, it may spread errors because it is a conditional two-qubit gate.
Below we list how local Pauli errors are spread by $U_{\mathrm{sz}}$:
\begin{align}\label{usz}
  U_{\mathrm{sz}}(\sigma_x\otimes I)U_{\mathrm{sz}}=\sigma_z\otimes\sigma_x, &
  U_{\mathrm{sz}}(I\otimes \sigma_x)U_{\mathrm{sz}}=\sigma_x\otimes\sigma_z; \nonumber\\
  U_{\mathrm{sz}}(\sigma_z\otimes I)U_{\mathrm{sz}}=I\otimes\sigma_z, &
  U_{\mathrm{sz}}(I\otimes \sigma_z)U_{\mathrm{sz}}=\sigma_z\otimes I;  \nonumber\\
  U_{\mathrm{sz}}(\sigma_y\otimes I)U_{\mathrm{sz}}=\sigma_z\otimes\sigma_y, &
  U_{\mathrm{sz}}(I\otimes \sigma_y)U_{\mathrm{sz}}=\sigma_y\otimes\sigma_z.
\end{align}
The spread relation for any other two-qubit Pauli error can be obtained from Eq. (\ref{usz}) by using the decomposition
\begin{equation}
U_{\mathrm{sz}}(\sigma_\alpha\otimes \sigma_\beta)U_{\mathrm{sz}}=U_{\mathrm{sz}}
(\sigma_\alpha\otimes I)U_{\mathrm{sz}} U_{\mathrm{sz}}(I\otimes \sigma_\beta)U_{\mathrm{sz}}.
\end{equation}
Here we model errors as the perfect application of a gate followed by one of the two-qubit Pauli errors
with probability $p/15$.
Our strategy to study the error spread is to examine how the Pauli errors caused by the auxiliary qubit spread
in a single holonomic CNOT gate in the first place, and then to trace them through a whole stabilizer circuit,
as shown in the main text.

\begin{figure}
  \centering
  \includegraphics[width=0.48\textwidth]{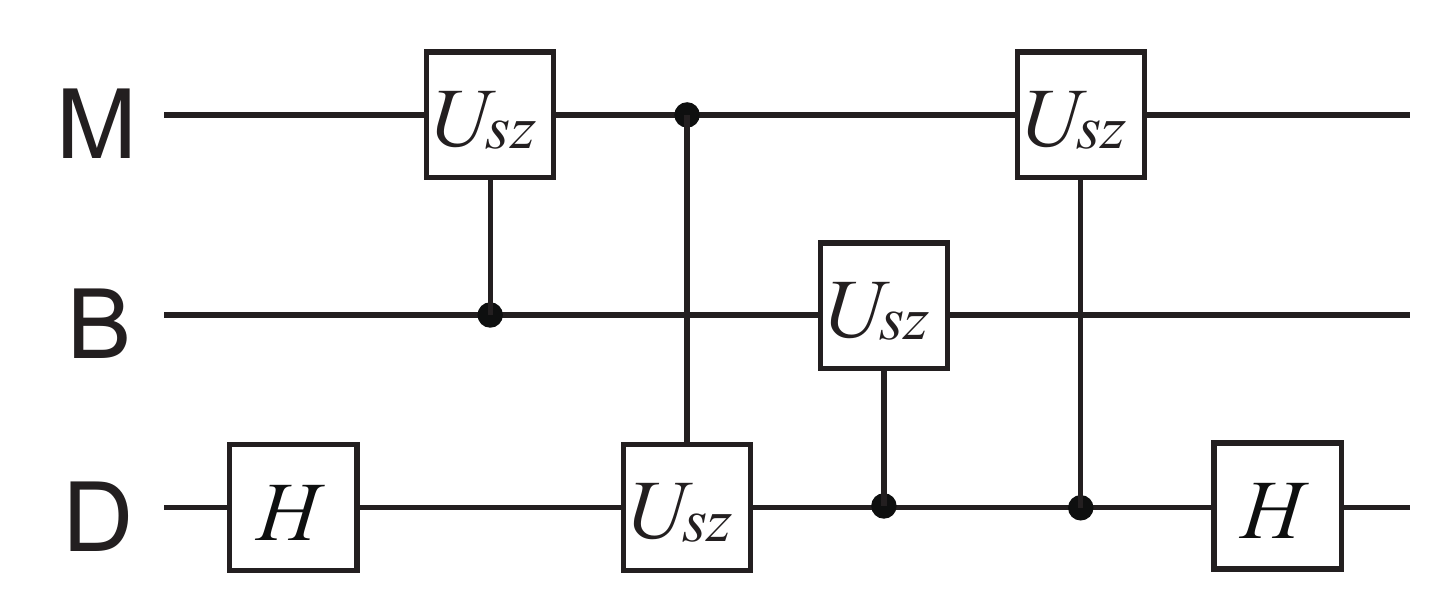}
  \caption{Quantum circuit for a holonomic CNOT gate acting on qubits $M$ and $D$, where qubit $B$ is initially prepared
  in $\ket{0}$. Here we choose qubit $D$ as the control qubit, so there are two Hadamard gates applied before and after
  the four $U_{\mathrm{sz}}$ gates. On the contrary, if we choose qubit $M$ as the control qubit, the two Hadamard
  gates should be applied to qubit $M$.}\label{hcc}
\end{figure}

To implement a holonomic CNOT gate on qubits $M$ and $D$, we need to perform the quantum circuit shown in Fig.~\ref{hcc}.
Below we examine how Pauli errors are spread through the circuit. For convenience, we write the initial state of the
qubits $M$, $B$, and $D$ as $\ket{\psi^M}\ket{0}\ket{\psi^D}$, where $\ket{\psi^{M(D)}}$ is the state of qubit $M(D)$.
Note that we write the state of qubits $M$ and $D$ in such a product state because an entangled state can be expressed
in a superposition of this form and it does not change our result. The state with error after a given gate operation
is listed below.\\

$Step 1$ ($H$).\\
$(\sigma_{\alpha1}^D)_D\ket{\psi^M}\ket{0}\ket{\psi^D_H}$, where
$\ket{\psi_H^D}=H\ket{\psi^D}$. Here the Pauli operator $\sigma_{\alpha1}^D$ in $(\dots)_D$ is the error followed
by implementing $H$ on qubit $D$. The superscript of $\sigma_{\alpha1}^D$ denotes the error which is originally
generated on qubit $D$ and the subscript $\alpha i$ implies that the error is the Pauli-$\alpha$ ($\alpha=x,y,z,0$)
operator generated in the $i$-th step (here, $\alpha=0$ corresponds to the identity operator $I$).
When we write the state of the three qubits, we always put the state of qubit $M$ in the first place and the state
of qubit $B$ at the second place, and so on.\\

$Step 2$ ($U_{\mathrm{sz}}^{MB}$). \\
$(\sigma_{\alpha2}^M)_M(\sigma_{\alpha2}^B)_B(\sigma_{\alpha1}^D)_D\ket{0}\ket{\psi^M}\ket{\psi^D_H}$. \\

$Step 3$ ($U_{\mathrm{sz}}^{DM}$).\\
$(\sigma_{\alpha3}^M\sigma_{\alpha1}^D\sigma_{2p}^M\sigma_{\alpha2}^M)_M(\sigma_{\alpha2}^B)_B(\sigma_{\alpha3}^D
\sigma_{\alpha2}^M\sigma_{2p}^D\sigma_{\alpha1}^D)_D\ket{\psi^D_H}\ket{\psi^M}\ket{0}$, where $\sigma_{2p}^i$
is an operator depending on $\sigma_{\alpha2}^i$ (when $\alpha2=x$ or $y$, $2p=z$; otherwise, $2p=0$).\\

$Step 4$ ($U_{\mathrm{sz}}^{BD}$).\\
$(\sigma_{\alpha3}^M\sigma_{\alpha1}^D\sigma_{2p}^M\sigma_{\alpha2}^M)_M
(\sigma_{\alpha4}^B\sigma_{\alpha3}^D\sigma_{\alpha2}^M\sigma_{2p}^D\sigma_{\alpha1}^D\sigma_{3p}^B\sigma_{\alpha2}^B)_B$\\
$(\sigma_{\alpha4}^D\sigma_{\alpha2}^B\sigma_{3p}^D\sigma_{\alpha3}
^D\sigma_{\alpha2}^M\sigma_{2p}^D\sigma_{\alpha1}^D)_D
\ket{\psi^D_H}\ket{0}\ket{\psi^M}$.\\

$Step 5$ ($U_{\mathrm{sz}}^{MD}$).\\
$(\sigma_{\alpha5}^M\sigma_{\alpha4}^D\sigma_{\alpha2}^B\sigma_{3p}^D\sigma_{\alpha3}^D\sigma_{\alpha2}^M\sigma_{2p}^D
\sigma_{\alpha1}^D\sigma_{4p}^M\sigma_{\alpha3}^M\sigma_{\alpha1}^D\sigma_{2p}^M\sigma_{\alpha2}^M)_M$\\
$(\sigma_{\alpha4}^B\sigma_{\alpha3}^D\sigma_{\alpha2}^M\sigma_{2p}^D\sigma_{\alpha1}^D\sigma_{3p}^B\sigma_{\alpha2}^B)_B$\\
$(\sigma_{\alpha5}^D\sigma_{\alpha3}^M\sigma_{\alpha1}
^D\sigma_{2p}^M\sigma_{\alpha2}^M\sigma_{4p}^D\sigma_{\alpha2}^B
\sigma_{3p}^D\sigma_{\alpha3}^D\sigma_{\alpha2}^M\sigma_{2p}^D\sigma_{\alpha1}^D)_D$\\
$U_{\mathrm{sz}}^{MD}\ket{\psi^D_H}\ket{0}\ket{\psi^M}$.\\

$Step 6$ ($H$).\\
$(\sigma_{\alpha5}^M\sigma_{\alpha4}^D\sigma_{\alpha2}^B\sigma_{3p}^D\sigma_{\alpha3}^D\sigma_{\alpha2}^M\sigma_{2p}^D
\sigma_{\alpha1}^D\sigma_{4p}^M\sigma_{\alpha3}^M\sigma_{\alpha1}^D\sigma_{2p}^M\sigma_{\alpha2}^M)_M$\\
$(\sigma_{\alpha4}^B\sigma_{\alpha3}^D\sigma_{\alpha2}^M\sigma_{2p}^D\sigma_{\alpha1}^D\sigma_{3p}^B\sigma_{\alpha2}^B)_B$\\
$(\sigma_{\alpha6}^D\sigma_{\alpha5}^D\sigma_{\alpha3}^M\sigma_{\alpha1}
^D\sigma_{2p}^M\sigma_{\alpha2}^M\sigma_{4p}^D\sigma_{\alpha2}^B
\sigma_{3p}^D\sigma_{\alpha3}^D\sigma_{\alpha2}^M\sigma_{2p}^D\sigma_{\alpha1}^D)_D$\\
$U_{\mathrm{sz}}^{MD}\ket{\psi^D_H}\ket{0}\ket{\psi^M_H}$.\\

After applying the gates in this circuit, we find that qubit $B$ is not entangled with qubits $M$ and $D$.
This implies that we can do a measurement on qubit $B$ without disturbing the other two qubits.
After the measurement, the Pauli errors
($\sigma_{\alpha4}^B\sigma_{\alpha3}^D\sigma_{\alpha2}^M\sigma_{2p}^D\sigma_{\alpha1}^D\sigma_{3p}^B\sigma_{\alpha2}^B$)
on qubit $B$ are eliminated so that it can be reset to $\ket{0}$.
On the other hand, since we use an auxiliary qubit $B$, the error $\sigma_{\alpha2}^B$ induced by qubit $B$
is spread to qubits $M$ and $D$ in a CNOT-gate circuit.

\begin{figure}[h]
  \centering
  \includegraphics[width=\columnwidth]{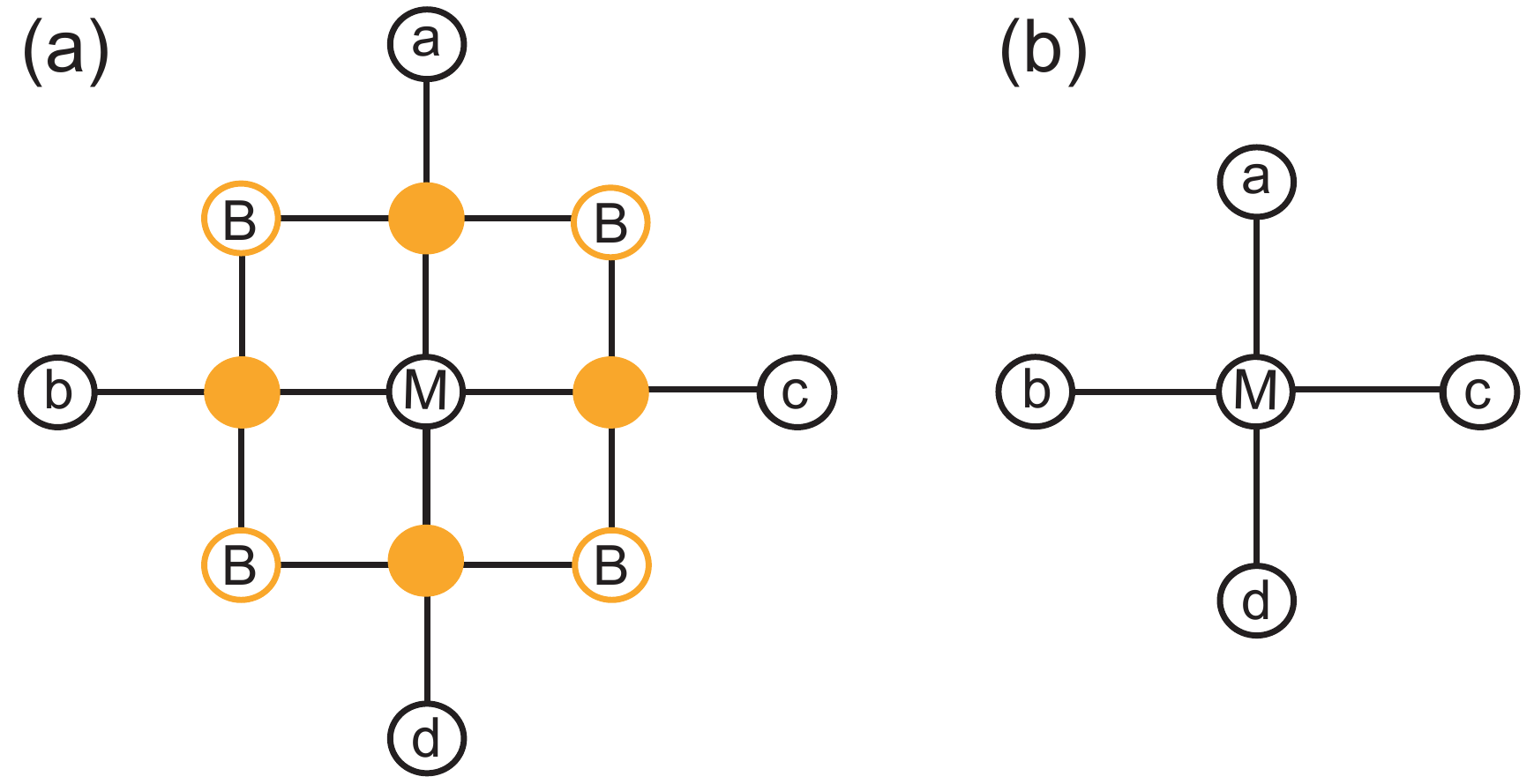}
  \caption{(a) The stabilizer unit for a measurement qubit in holonomic surface codes, where the four data qubits
  are denoted by $a$, $b$, $c$, and $d$. Here qubit $M$ is the measurement qubit and each qubit $B$ is
  an auxiliary qubit. (b) The stabilizer unit for a measurement qubit in dynamic surface codes. }\label{hdep}
\end{figure}

Now we check how the error $\sigma_{\alpha2}^B$ spreads through the whole stabilizer circuit. For simplicity,
we omit the Pauli errors induced by qubits $M$ and $D$ because they also occur in the circuit for dynamic surface codes.
For the holonomic stabilizer unit shown in Fig. \ref{hdep}(a), we list the corresponding error-spread path in Table I.

\begin{table*}\label{table1}\caption{This table shows how the error $\sigma_{\alpha2}^{B}(i)$ spreads through
a stabilizer circuit, where $\sigma_{\alpha2}^{B}(i)$ ($i=a,b,c,d$) is the error occurring in the CNOT ($M$,$i$)
gate between the measurement qubit $M$ and the $i$th data qubit. The four CNOT gates in the left column are
implemented successively in the stabilizer circuit and the corresponding accumulated errors at both the measurement
qubit $M$ and the $i$th data qubit are listed in the other columns.}
  \centering
  \begin{tabular}{cccccc}
  \hline\hline
   & & &Qubits & & \\
  Operations & $M$ & $a$ & $b$ & $c$ & $d$ \\ \hline
  CNOT ($M$,$a$) & $\sigma_{\alpha2}^{B}(a)$ & $\sigma_{\alpha2}^{B}(a)$ & $I$ & $I$ & $I$ \\ \hline
  CNOT ($M$,$b$) & $\sigma_{\alpha2}^{B}(b)\sigma_{\alpha2}^{B}(a)$ & $\sigma_{\alpha2}^{B}(a)$ &
  $\sigma_{\alpha2}^{B}(a)$ & $I$ & $I$ \\ \hline
  CNOT ($M$,$c$) & $\sigma_{\alpha2}^{B}(c)\sigma_{\alpha2}^{B}(b)\sigma_{\alpha2}^{B}(a)$ &
  $\sigma_{\alpha2}^{B}(a)$ & $\sigma_{\alpha2}^{B}(b)\sigma_{\alpha2}^{B}(a)$ &
  $\sigma_{\alpha2}^{B}(c)\sigma_{\alpha2}^{B}(b)\sigma_{\alpha2}^{B}(a)$ & $I$ \\ \hline
  CNOT ($M$,$d$) & $\sigma_{\alpha2}^{B}(d)\sigma_{\alpha2}^{B}(c)\sigma_{\alpha2}^{B}(b)\sigma_{\alpha2}^{B}(a)$
  & $\sigma_{\alpha2}^{B}(a)$ & $\sigma_{\alpha2}^{B}(b)\sigma_{\alpha2}^{B}(a)$ &
  $\sigma_{\alpha2}^{B}(c)\sigma_{\alpha2}^{B}(b)\sigma_{\alpha2}^{B}(a)$ &
  $\sigma_{\alpha2}^{B}(d)\sigma_{\alpha2}^{B}(c)\sigma_{\alpha2}^{B}(b)\sigma_{\alpha2}^{B}(a)$ \\
  \hline
\end{tabular}
\end{table*}

The possibility to form an undetectable error ($d=4$) with the Pauli errors induced by four auxiliary qubits $B$'s
can be calculated as
$\frac{12}{15}\times\frac{3}{15}\times\frac{3}{15}\times\frac{3}{15}=\frac{4}{625}$, which is quite small.
Also, it is easy to check that the error-spread path just mimics that for dynamic surface codes.
Therefore, the implementation of the holonomic stabilizer circuit does not spread Pauli errors worse than the
implementation of the dynamic one.

\section{Robustness to stochastic fluctuations}

In this section, we show that holonomic quantum gates are robust against small stochastic fluctuations.
Let us study a quantum system described by a Hamiltonian $\hat{H}$ which has eigenstates $\{\ket{\psi_k}\}$.
When the considered computational subspace experiences a nonadiabatic cyclic evolution in the total Hilbert
space of $\hat{H}$, its evolution is described by~\cite{sjoqvist12,xu12}
\begin{equation}\label{}
  U(t,0)=\mathrm{T}e^{-i\int_{0}^{t}[G(s)+D(s)]\mathrm{d}s},
\end{equation}
where $G(t)\equiv(G_{kl}(t))$ and $D(t)\equiv(D_{kl}(t))$, with
\begin{eqnarray}
G_{kl}(t)&=&-i\bra{\psi_k(t)}\frac{\partial}{\partial t}\ket{\psi_l(t)}, \nonumber\\
D_{kl}(t)&=&\bra{\psi_k(t)}\hat{H}\ket{\psi_l(t)},
\end{eqnarray}
are owing to the geometric and dynamic contributions, respectively.
In the ideal holonomic scheme, $\hat{H}=\hat{H}_1$ and $\hat{H}_2$ for the single- and two-qubit gates, which are
properly designed to have zero dynamic contribution $D_{kl}(t)=0$.
Below we consider the case with stochastic fluctuations occurring in the Hamiltonian of the system, i.e.,
$\tilde{H}_i(t)=\hat{H}_i+\delta \hat{H}^\prime_i(t)$ ($i=1,2$), where the small perturbation
$\delta\hat{H}^\prime_i(t)$ contains the stochastic fluctuations, and show that the holonomic gates are robust
against the small stochastic fluctuations.

(i) For the special case with $\delta\hat{H}^\prime_i(t)=\delta(t)\hat{H}_i$, there is
$\tilde{H}_i(t)=[1+\delta(t)]\hat{H}_i$. The corresponding dynamic contribution remains equal to zero because
\begin{align}
D_{kl}(t)=&\bra{\psi_k(t)}\tilde{H}_i(t)\ket{\psi_l(t)}\nonumber \\
=&[1+\delta(t)]\bra{\psi_k(t)}\hat{H}_i\ket{\psi_i(t)}=0,
\end{align}
indicating that the stochastic fluctuations do not give rise to dynamic contributions, and the system evolves
fully quantum holonomically.
Meanwhile, the fluctuations do make the real cyclic path deviate a bit from the ideal cyclic path because
\begin{align}
G_{kl}(t)&=-i\bra{\psi_k(t)}\frac{\partial}{\partial t}\ket{\psi_l(t)} \nonumber \\
&=-\bra{\psi_k}\tilde{U}^\dag(t)\tilde{H}(t)\tilde{U}(t)\ket{\psi_l},
\end{align}
where
\begin{equation}
\tilde{U}(t)=\mathrm{T}e^{-i\int_{0}^{t}\tilde{H}(s)\mathrm{d}s}.
\end{equation}
However, owing to the small stochastic fluctuations, the deviation along the whole cyclic path is averaged
by the path integration. Therefore, the net influence of the fluctuations can be ignored if the time used
for the cyclic evolution is long.

(ii) The general case is $[\delta\hat{H}^\prime_i(t),\hat{H}_i]\neq0$. In this case, $D_{kl}(t)$ does not
vanish because the qubit subspace is no longer a dark-invariant subspace for $\tilde{H}_i$. However,
for small stochastic fluctuations, the deviation of $D_{kl}(t)$ from 0 is also small. Then, we can write the
evolution operator as
\begin{align}\label{}
  U(t,0)=&\prod_{n=1}^{\infty}e^{-i\int_{t_n}^{t_{n+1}}(G(s)+D(s))\mathrm{d}s} \nonumber\\
        =&\prod_{n=1}^{\infty}e^{-i\int_{t_n}^{t_{n+1}}G(s)\mathrm{d}s}e^{-i\int_{t_n}^{t_{n+1}}D(s)
        \mathrm{d}s}\nonumber \\
        &e^{-\frac{i}{2}\int_{t_n}^{t_{n+1}}\int_{t_n}^{t_{n+1}}[G(s),D(s^{\prime})]\mathrm{d}s\mathrm{d}s^{\prime}+\cdots},
\end{align}
where we have used the Baker-Campbell-Hausdorff formula \cite{bch}. When the stochastic fluctuations are small enough,
$-\frac{i}{2}\int_{t_n}^{t_{n+1}}\int_{t_n}^{t_{n+1}}[G(s),D(s^{\prime})]\mathrm{d}s\mathrm{d}s^{\prime}+\cdots$
is a higher-order small term and $U(t,0)$ can be approximately written as
\begin{equation}\label{}
  U(t,0)\approx \prod_{n=1}^{\infty}e^{-i\int_{t_n}^{t_{n+1}}G(s)\mathrm{d}s}e^{-i\int_{t_n}^{t_{n+1}}D(s)\mathrm{d}s}.
\end{equation}
Now we are going to sort out the geometric evolutions from the dynamic ones.
For any two adjoining sections $n=j$ and $n=j+1$,
\begin{align}\label{}
  &e^{-i\int_{t_j}^{t_{j+1}}G(s)\mathrm{d}s}e^{-i\int_{t_j}^{t_{j+1}}D(s)\mathrm{d}s}
  e^{-i\int_{t_{j+1}}^{t_{j+2}}G(s)\mathrm{d}s} e^{-i\int_{t_{j+1}}^{t_{j+2}}D(s)\mathrm{d}s}  \nonumber \\
  &= e^{-i\int_{t_j}^{t_{j+1}}G(s)\mathrm{d}s}e^{-i\int_{t_{j+1}}^{t_{j+2}}G(s)\mathrm{d}s}
  e^{-i\int_{t_j}^{t_{j+1}}D(s)\mathrm{d}s}\nonumber \\
  &~~~ \times e^{-i\int_{t_{j+1}}^{t_{j+2}}
  \int_{t_{j}}^{t_{j+1}}[G(s),D(s^{\prime})]\mathrm{d}s\mathrm{d}s^{\prime}
  +\cdots}
  e^{-i\int_{t_{j+1}}^{t_{j+2}}D(s)\mathrm{d}s} \nonumber\\
  &\approx  e^{-i\int_{t_j}^{t_{j+1}}G(s)\mathrm{d}s}e^{-i\int_{t_{j+1}}^{t_{j+2}}G(s)\mathrm{d}s}
  e^{-i\int_{t_j}^{t_{j+1}}D(s)\mathrm{d}s}\nonumber \\
  &~~~ \times e^{-i\int_{t_{j+1}}^{t_{j+2}}D(s)\mathrm{d}s},
\end{align}
where we have again ignored the higher-order small term. When the stochastic fluctuations are small enough,
following the same procedure, we can approximately write the evolution operator as
\begin{align}
  U(t,0)&=\prod_{n=1}^{\infty}e^{-i\int_{t_n}^{t_{n+1}}G(s)\mathrm{d}s}e^{-i\int_{t_n}^{t_{n+1}}D(s)\mathrm{d}s}\nonumber \\
  &\approx\prod_{n=1}^{\infty}e^{-i\int_{t_n}^{t_{n+1}}G(s)\mathrm{d}s}\prod_{n=1}^{\infty}e^{-i\int_{t_n}^{t_{n+1}}D(s)
  \mathrm{d}s} \nonumber\\
  &=\mathrm{P}e^{-i\oint G(s)\mathrm{d}s}\mathrm{T}e^{-i\int_{0}^{t}D(s)\mathrm{d}s},
\end{align}
where $P$ denotes the path-integral ordering. While $D_{kl}(t)=0$ in the ideal holonomic scheme, the real dynamic
contributions $D_{kl}(t)$ become
\begin{align}
D_{kl}(t)=&\bra{\psi_k(t)}[\hat{H}_i+\delta \hat{H}^\prime_i(t)]\ket{\psi_l(t)}\nonumber \\
=&\bra{\psi_k(t)}\delta \hat{H}^\prime_i(t)\ket{\psi_l(t)}=\bra{\psi_k}\delta \hat{A}(t)\ket{\psi_l},
\end{align}
with
\begin{equation}
\delta \hat{A}(t)=\tilde{U}^\dag(t)\delta \hat{H}^\prime_i(t)\tilde{U}(t).
\end{equation}
Because $\delta\hat{H}^\prime_i(t)$ fluctuates stochastically around zero, $D_{kl}(t)$ behave like errors which also
fluctuate stochastically around zero.  Therefore, the dynamical evolution $\mathrm{T}e^{-i\int_{0}^{t}D(s)\mathrm{d}s}$
is related to the average of stochastic errors over time, which goes to zero (i.e., the dynamical evolution is reduced
to an identity operator) for a long time interval used for the whole cyclic path. For the  geometric evolution
$\mathrm{P}e^{-i\oint G(s)\mathrm{d}s}$, owing to the stochastic fluctuation $\delta\hat{H}^\prime_i(t)$, the real path
stochastically fluctuates around the ideal loop with $\delta\hat{H}^\prime_i(t)=0$.
However, the deviation from the ideal geometric evolution is averaged by the path integration along the whole
cyclic path, which also goes to zero when using a long-time interval for the cyclic evolution.

To explicitly show that the stochastic Pauli errors considered in the previous section can be regarded as
special cases of the stochastic fluctuations, let us consider, for example,
the Hamiltonian in Eq.~(\ref{h1}), i.e.,
\begin{equation}
  \hat{H}_1=\frac{J_1}{2}(\cos\beta\sigma_x^1+\sin\beta\sigma_y^1)
+\frac{J_{12}}{2}(\sigma_x^1\sigma_x^2+\sigma_y^1\sigma_y^2).
\end{equation}
When stochastic fluctuations occur, $\hat{H}_1$ becomes $\tilde{H}_1(t)=\hat{H}_1+\delta \hat{H}^\prime_1(t)$,
where the fluctuation term can be generally written as
\begin{eqnarray}
\delta\hat{H}_1^\prime(t)&=&\delta_1^{(x)}(t)\sigma_x^1+\delta_1^{(y)}(t)\sigma_y^1
+\delta_{12}^{(x)}(t)\sigma_x^1\sigma_x^2 \nonumber\\
&&+\delta_{12}^{(y)}(t)\sigma_y^1\sigma_y^2.
\end{eqnarray}
The evolution operator of the system is
\begin{equation}\label{}
  \tilde{U}_1(t)=\mathrm{T}e^{-i\int_{0}^{t}[\hat{H}_1(s)+\delta \hat{H}^\prime_1(s)]\mathrm{d}s},
\end{equation}
When the stochastic fluctuation $\delta \hat{H}^\prime_1(t)$ is small enough, we can write the evolution operator as
\begin{eqnarray}\label{ue}
\tilde{U}_1(t)&\approx & \mathrm{T}e^{-i\int_{0}^{t}\hat{H}_1(s)\mathrm{d}s}\mathrm{T}e^{-i\int_{0}^{t}\delta
\hat{H}^\prime_1(s)\mathrm{d}s} \nonumber \\
&\approx & \mathrm{T}e^{-i\int_{0}^{t}\hat{H}_1(s)\mathrm{d}s}[1-i\int_{0}^{t}\delta\hat{H}^\prime(s)\mathrm{d}s].
\end{eqnarray}
The first term in Eq.~(\ref{ue}) corresponds to the ideal unitary evolution of the system
$U_1(t)=\mathrm{T}e^{-i\int_{0}^{t}\hat{H}_1(s)\mathrm{d}s}$, and the second term corresponds to the errors due to
$\delta\hat{H}^\prime(t)$.
For instance, when $U_1(t)=\sigma_x^1$, the second term in Eq.~(\ref{ue}) becomes
\begin{eqnarray}\label{ue2}
  -i\sigma_x^1\int_{0}^{t}\delta\hat{H}^\prime(s)\mathrm{d}s &=&-i\int_{0}^{t}\delta_1^{(x)}(s)\mathrm{d}s+
  \sigma_z^1\int_{0}^{t}\delta_1^{(y)}(s)\mathrm{d}s \nonumber \\
  & &-i\sigma_x^2\int_{0}^{t}\delta_{12}^{(x)}(s)\mathrm{d}s \nonumber  \\
  & &+\sigma_z^1\sigma_y^2\int_{0}^{t}\delta_{12}^{(y)}(s)\mathrm{d}s.
\end{eqnarray}
It gives rise to the Pauli error $\sigma_z\otimes I$ on qubit 1, the Pauli error $I\otimes\sigma_x$ on qubit 2,
and the Pauli error $\sigma_z\otimes\sigma_y$ on both qubit 1 and qubit 2.
When higher-order terms in the expansion of
$\mathrm{T}e^{-i\int_{0}^{t}\delta \hat{H}^\prime_1(s)\mathrm{d}s}$ are included, each of them, together with
$U_1(t)=\sigma_x^1$,
will also give rise to the Pauli errors of the types $\sigma_{\alpha}\otimes I$, $I\otimes\sigma_{\beta}$,
and $\sigma_{\alpha}\otimes\sigma_{\beta}$, where $\alpha,\beta=x$, $y$, or $z$.

\section{Fidelities of the holonomic and dynamic two-qubit gates}

As explained above, holonomic quantum gates are robust against small stochastic fluctuations in the Hamiltonian
of the system. This is a distinct advantage of the quantum-holonomy scheme. Below we further consider the area error
due to the imperfect control of the parameters in the Hamiltonian, because it may now become the main source of errors
in the holonomic quantum gates.

\subsection{Dynamic two-qubit gate}

For a dynamic two-qubit iSWAP gate built on a pair of measurement and data qubits coupled directly via an $XY$-type
interaction, the Hamiltonian reads
\begin{equation}\label{}
  \hat{H}_d(t)=\frac{\Omega(t)}{2}(\sigma_x^M\sigma_x^D+\sigma_y^M\sigma_y^D).
\end{equation}
By turning on $\hat{H}_d$ for a time $\tau$ so that $\int_{0}^{\tau}\Omega(s)\mathrm{d}s=\frac{3\pi}{2}$,
the corresponding time-evolution operator turns out to be an iSWAP gate,
\begin{equation}\label{iSWAP}
  U_{\mathrm{iSWAP}}=e^{-i\int_{0}^{\tau}\hat{H}_d(t)\mathrm{d}t}=\left(
                                           \begin{array}{cccc}
                                             1 & 0 & 0 & 0 \\
                                             0 & 0 & i & 0 \\
                                             0 & i & 0 & 0 \\
                                             0 & 0 & 0 & 1 \\
                                           \end{array}
                                         \right),
\end{equation}
in the ordered basis $\{\ket{00},\ket{01},\ket{10},\ket{11}\}$.
In the presence of an area error $\delta$, we replace
$\int_{0}^{\tau}\Omega(s)\mathrm{d}s=\frac{3\pi}{2}$ with $\int_{0}^{\tau^\prime}\Omega(s)\mathrm{d}s
=\frac{3\pi}{2}+\delta$. With respect to the iSWAP gate in Eq.~(\ref{iSWAP}), the corresponding gate with this area
error can be written as
\begin{equation}\label{dyna}
  U_d(\tau^\prime)=\left(
                     \begin{array}{cccc}
                       1 & 0 & 0 & 0 \\
                       0 & \sin\delta & i\cos\delta & 0 \\
                       0 & i\cos\delta & \sin\delta & 0 \\
                       0 & 0 & 0 & 1 \\
                     \end{array}
                   \right).
\end{equation}

For a given initial state $ \ket{\psi}_i=a_{00}\ket{00}+a_{01}\ket{01}+a_{10}\ket{10}+a_{11}\ket{11}$,
where $a_{00}, a_{01}, a_{10}$, and $a_{11}$ are complex numbers that satisfy the normalization condition,
when the iSWAP gate in Eq.~(\ref{iSWAP}) is applied, the ideal final state can be obtained as
\begin{equation}\label{}
  \ket{\psi^{(i)}}_f=a_{01}\ket{00}+ia_{10}\ket{01}+ia_{01}\ket{10}+a_{11}\ket{11}.
\end{equation}
Given the same initial state, when the corresponding gate in Eq.~(\ref{dyna}) is applied, the real final state is given by
\begin{align}
  \ket{\psi^{(r)}}_f= & a_{00}\ket{00}+(a_{01}\sin\delta+ia_{10}\cos\delta)\ket{01}\nonumber \\
  &+(ia_{01}\cos\delta+a_{10}\sin\delta )\ket{10}+a_{11}\ket{11}.
\end{align}
Therefore, the fidelity between the ideal and real final states can be obtained as
\begin{align}
  F\equiv &|\braket{\psi^{(i)}}{\psi^{(r)}}_f|=\large|1-2\sin^2\frac{\delta}{2}(|a_{01}|^2+|a_{10}|^2) \nonumber \\
  &-i\sin\delta(a_{01}^\ast a_{10}+a_{01}a_{10}^\ast)|.
\end{align}
Note that $|a_{01}|^2+|a_{10}|^2\in[0,1]$. When $a_{01}=1$ and $a_{10}=0$, the fidelity reaches its lower bound that reads
\begin{equation}\label{}
  F_d=1-\frac{\delta^2}{2},
\end{equation}
up to second order in $\sin\delta$.

\subsection{Holonomic single-qubit gate}

In our approach, a holonomic single-qubit gate involves two physical qubits, with one designed as the target qubit
and the other as the auxiliary qubit. Thus, its fidelity can be compared directly with that of a dynamic two-qubit gate.
As shown in the main text, one needs to turn $\hat{H}_1$ on for some time $t$ to realize the holonomic single-qubit gate,
so that the cyclic condition can be satisfied, i.e.,
$a_t\cos^2\frac{\theta}{2}=(2n+1)\pi$ and $a_t\sin^2\frac{\theta}{2}=2n\pi$. Also, the auxiliary qubit
should be initially prepared in the state $\ket{0}$.
In the presence of an area error $\delta$, the real cyclic condition becomes
$a^\prime_t\cos^2\frac{\theta}{2}=(2n+1)\pi+\delta\cos^2\frac{\theta}{2}$ and
$a^\prime_t\sin^2\frac{\theta}{2}=2n\pi+\delta\sin^2\frac{\theta}{2}$.
The matrix elements in the evolution operator $U_1$ [see Eq.~(3)] are now given by
\begin{widetext}
\begin{align}\label{}
  &V_0\cos(a^\prime_tD)V_0^\dag=
  \left(
  \begin{array}{cc}
  -\sin^2\frac{\theta}{2}\cos(\delta\cos^2\frac{\theta}{2})+\cos^2\frac{\theta}{2}\cos(\delta\sin^2\frac{\theta}{2})
  & -\frac{1}{2}e^{-i\beta}\sin\theta[\cos(\delta\cos^2\frac{\theta}{2})+\cos(\delta\sin^2\frac{\theta}{2})] \\
  -\frac{1}{2}e^{i\beta}\sin\theta[\cos(\delta\cos^2\frac{\theta}{2})+\cos(\delta\sin^2\frac{\theta}{2})]  &
  -\cos^2\frac{\theta}{2}\cos(\delta\cos^2\frac{\theta}{2})+\sin^2\frac{\theta}{2}\cos(\delta\sin^2\frac{\theta}{2}) \\
                                 \end{array}
                               \right), \nonumber\\
  &V_0\sin(a^\prime_tD)V_1^\dag=\left(
                                  \begin{array}{cc}
  \frac{1}{2}\sin\theta[\sin(\delta\sin^2\frac{\theta}{2})-\sin(\delta\cos^2\frac{\theta}{2})] &
  e^{-i\beta}[\sin^2\frac{\theta}{2}\sin(\delta\cos^2\frac{\theta}{2})+\cos^2\frac{\theta}{2}
  \sin(\delta\sin^2\frac{\theta}{2})] \nonumber \\
  -e^{i\beta}[\sin^2\frac{\theta}{2}\sin(\delta\sin^2\frac{\theta}{2})+\cos^2\frac{\theta}{2}
  \sin(\delta\cos^2\frac{\theta}{2})]  &
  \frac{1}{2}\sin\theta[-\sin(\delta\cos^2\frac{\theta}{2})+\sin(\delta\sin^2\frac{\theta}{2})] \\
                                  \end{array}
                                \right), \\
  &V_1\cos(a^\prime_tD)V_1^\dag=\left(
                                  \begin{array}{cc}
  -\cos^2\frac{\theta}{2}\cos(\delta\cos^2\frac{\theta}{2})+\sin^2\frac{\theta}{2}\cos(\delta\sin^2\frac{\theta}{2}) &
  -\frac{1}{2}e^{-i\beta}\sin\theta[\cos(\delta\cos^2\frac{\theta}{2})+\cos(\delta\sin^2\frac{\theta}{2})] \\
  -\frac{1}{2}e^{i\beta}\sin\theta[\cos(\delta\cos^2\frac{\theta}{2})+\cos(\delta\sin^2\frac{\theta}{2})] &
  -\sin^2\frac{\theta}{2}\cos(\delta\cos^2\frac{\theta}{2})+\cos^2\frac{\theta}{2}\cos(\delta\sin^2\frac{\theta}{2})\\
                                  \end{array}
                                \right).
\end{align}
\end{widetext}
Assume that the initial state of the two physical qubits is $(a\ket{0}+b\ket{1})\otimes\ket{0}$,
where $|a|^2+|b|^2=1$ and the auxiliary qubit is in the ground state $\ket{0}$. With $U_1$ applied to these
two qubits, the real final state reads
\begin{widetext}
\begin{align}\label{}
  \ket{\psi^{(r)}_1}=&\left\{a\left[-\sin^2\frac{\theta}{2}\cos\left(\delta\cos^2\frac{\theta}{2}\right)
  +\cos^2\frac{\theta}{2}\cos\left(\delta\sin^2\frac{\theta}{2}\right)\right]
  -\frac{b}{2}e^{-i\beta}\sin\theta\left[\cos\left(\delta\cos^2\frac{\theta}{2}\right)
  +\cos\left(\delta\sin^2\frac{\theta}{2}\right)\right]\right\}\ket{00} \nonumber \\
  &-\left\{\frac{a}{2}e^{i\beta}\sin\theta\left[\cos\left(\delta\cos^2\frac{\theta}{2}\right)
  +\cos\left(\delta\sin^2\frac{\theta}{2}\right)\right]
  -b\left[\cos^2\frac{\theta}{2}\cos\left(\delta\cos^2\frac{\theta}{2}\right)
  -\sin^2\frac{\theta}{2}\cos\left(\delta\sin^2\frac{\theta}{2}\right)\right]\right\}\ket{01} \nonumber \\
  &-i\left\{\frac{a}{2}\sin\theta\left[\sin\left(\delta\sin^2\frac{\theta}{2}\right)
  -\sin\left(\delta\cos^2\frac{\theta}{2}\right)\right]
  -be^{-i\beta}\left[\sin^2\frac{\theta}{2}\sin\left(\delta\sin^2\frac{\theta}{2}\right)
  +\cos^2\frac{\theta}{2}\sin\left(\delta\cos^2\frac{\theta}{2}\right)\right]\right\}\ket{10} \nonumber \\
  &-i\left\{ae^{i\beta}\left[\sin^2\frac{\theta}{2}\sin\left(\delta\cos^2\frac{\theta}{2}\right)
  +\cos^2\frac{\theta}{2}\sin\left(\delta\sin^2\frac{\theta}{2}\right)\right] +\frac{b}{2}\sin\theta
  \left[-\sin\left(\delta\cos^2\frac{\theta}{2}\right)
  +\sin\left(\delta\sin^2\frac{\theta}{2}\right)\right]\right\}\ket{11}.
\end{align}
\end{widetext}
Without the area error ($\delta=0$), the ideal final state is
\begin{equation}
  \ket{\psi^{(i)}_1}=(a\cos\theta-be^{-i\beta}\sin\theta)\ket{00}-(ae^{i\beta}\sin\theta+b\cos\theta)\ket{01}.
\end{equation}
For the holonomic Hadamard gate, $\theta=\frac{\pi}{4}$ and $\beta=0$ .
The fidelity of this holonomic single-qubit gate can be obtained as
\begin{equation}
  F\approx1-\frac{\delta^4}{32}(1-|ab|).
\end{equation}
In particular, when $|a|=|b|=\frac{1}{\sqrt{2}}$, the fidelity reaches its lower bound
\begin{equation}
F_h=1-\frac{\delta^4}{64},
\end{equation}
which is much better than the fidelity of the holonomic two-qubit gate given below [see Eq.~(\ref{hfide1})].

\subsection{Holonomic two-qubit gate}

In contrast to the dynamic iSWAP gate, when the holonomic $U_{\mathrm{sz}}$ gate in Eq.~(\ref{E4}) is applied,
the initial state $\ket{\psi}_i\otimes\ket{0}$ of the three qubits, where $\ket{0}$ is the state of the auxiliary qubit,
is transformed to
\begin{equation}\label{}
  \ket{\psi^{(i)}}_f=(a_{00}\ket{00}+a_{10}\ket{01}+a_{01}\ket{10}-a_{11}\ket{11})\otimes\ket{0}.
\label{ideal}
\end{equation}
When including the area error $\delta$, the cyclic condition $\int_0^\tau\Omega(s)\mathrm{d}s=\pi$ becomes
$\int_0^{\tau^\prime}\Omega(s)\mathrm{d}s=\pi+\delta$.
Given the same initial state $\ket{\psi}_i\otimes\ket{0}$, the real final state of the three qubits can be obtained as
\begin{equation}
  \ket{\psi^{(r)}}_f=\ket{\psi}_0\otimes\ket{0}+i\sin\delta\ket{\psi}_1\otimes\ket{1},
\label{final}
\end{equation}
where
\begin{align}
\ket{\psi}_0=&a_{00}\ket{00}+\left(a_{01}\sin^2\frac{\delta}{2}+a_{10}\cos^2\frac{\delta}{2}\right)\ket{01}\nonumber\\
&+\left(a_{10}\sin^2\frac{\delta}{2}+a_{01}\cos^2\frac{\delta}{2}\right)\ket{10}-a_{11}\cos\delta\ket{11},\nonumber\\
\ket{\psi}_1=&\frac{\sqrt{2}}{2}(a_{01}+a_{10})\ket{00}+\frac{\sqrt{2}}{2}a_{11}\ket{10}+\frac{\sqrt{2}}{2}a_{11}\ket{01}.
\end{align}
Since $\delta$ is a small quantity, we replace $\cos\delta$ with $1-\delta^2/2$
and $\sin\delta$ with $\delta$ (up to second order). Thus, the fidelity of the considered holonomic two-qubit gate can
be written as
\begin{equation}\label{s19}
  F=\frac{1-\alpha\delta^2}{1-\alpha\delta^2+\alpha\delta^4},
\end{equation}
where
\begin{equation}
\alpha=\frac{1}{2}(|a_{01}|^2+|a_{10}|^2+a_{01}^\ast a_{10}+a_{01}a_{10}^\ast).
\end{equation}
It is clear that $F$ decreases when increasing $\alpha$. Also, it is important to note that $\alpha\in[0,1]$
because $a_{01}$ and $a_{10}$ must satisfy the normalization condition. Therefore, the lower bound of the fidelity
for the holonomic gate $U_{\mathrm{sz}}$ is given by
\begin{equation}\label{hfide1}
  F_h=1-\frac{\delta^4}{1-\delta^2+\delta^4}\approx 1-\delta^4,
\end{equation}
which corresponds to $\alpha=1$ ($a_{01}=a_{10}=\frac{\sqrt{2}}{2}$) in Eq.~(\ref{s19}).

\section{Two different measurements on the auxiliary qubit}

\subsection{Measurement with $\sigma_z$}

Owing to the area error $\delta$, the two target qubits become entangled with the auxiliary qubit in the real final state
$\ket{\psi^{(r)}}_f$ given in Eq.~(\ref{final}). Therefore, we need to implement a measurement on the auxiliary qubit,
so as to reset the  auxiliary qubit to its ground state $\ket{0}$. If we implement this measurement with $\sigma_z$,
there are two possible outcomes. When $\ket{0}$ is obtained for the auxiliary qubit, the state of the two target
qubits is collapsed to $\ket{\psi}_0$ with a very high probability, while the state of the two target qubits is
collapsed to $\ket{\psi}_1$ with a very low probability when $\ket{1}$ is obtained for the auxiliary qubit.
Although $\ket{\psi}_0\otimes\ket{0}$ deviates a bit from the ideal final state $\ket{\psi^{(i)}}_f$,
it is perfectly correctable by surface codes, as long as $\delta$ is small enough. Here, $\ket{\psi}_1\otimes\ket{1}$
is free of the area error $\delta$, but we need to have it recovered to $\ket{\psi^{(i)}}_f$.

\begin{figure}[h]
  \centering
  \includegraphics[width=0.47\textwidth]{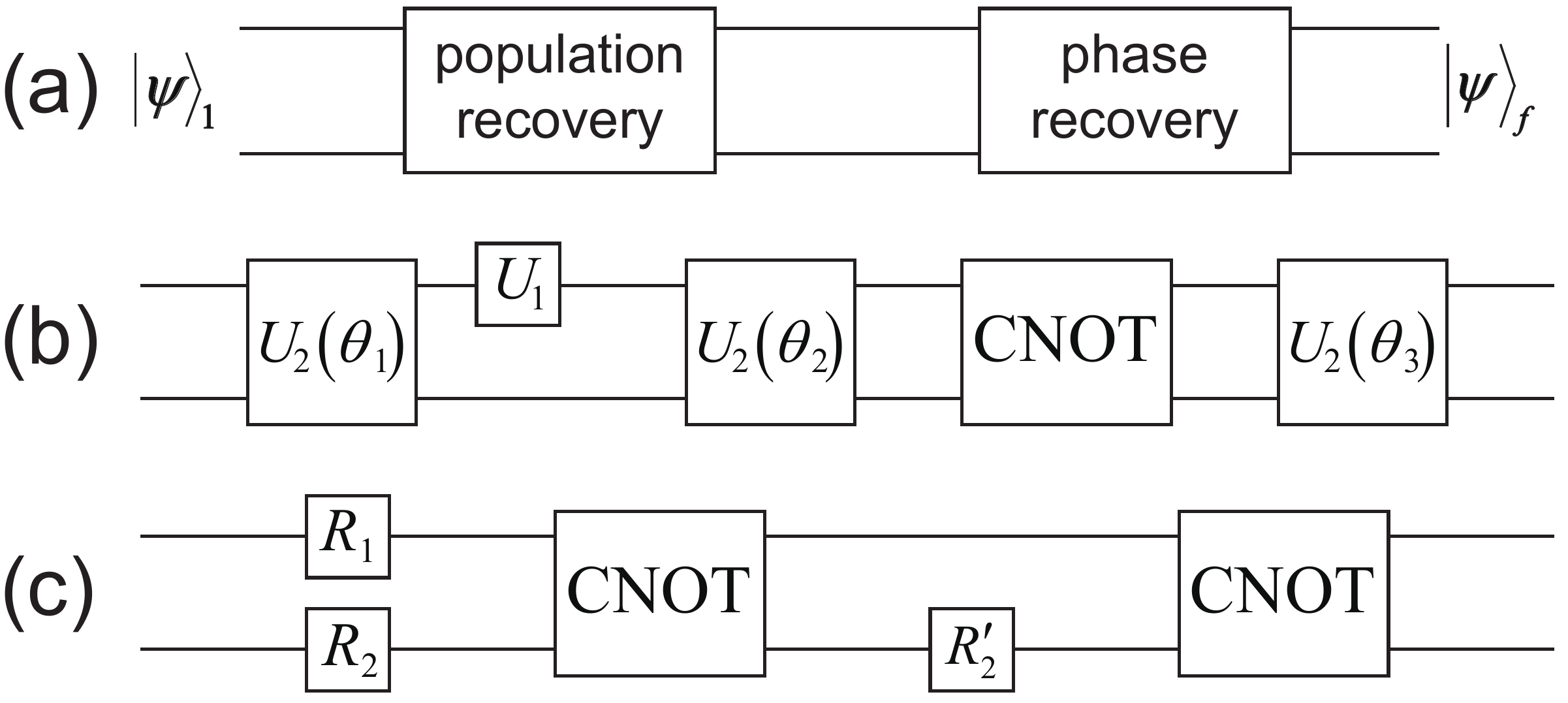}
  \caption{Quantum circuit for the state recovery of $\ket{\psi}_1$ to
  $\ket{\psi}_f= a_{00}\ket{00}+a_{10}\ket{01}+a_{01}\ket{10}-a_{11}\ket{11}$.
  (a) The recovery is divided into two steps, i.e., the  population recovery and the phase recovery,
  both of which are also accomplished with holonomic gates. (b) Explicit circuit for the population recovery.
  (c) Explicit circuit for the phase recovery. }
  \label{recovery}
\end{figure}

For convenience, we write the coefficients $a_{ij}$ of the initial state
$\ket{\psi}_i\equiv a_{00}\ket{00}+a_{01}\ket{01}+a_{10}\ket{10}+a_{11}\ket{11})$ as $|a_{ij}|e^{i\phi_{ij}}$.
Our strategy is to complete the state recovery task for $\ket{\psi}_1$ in two steps (see Fig. \ref{recovery}):
First, we recover the population of each basis $\ket{ij}$, so as to transfer $\ket{\psi}_1$ to
$\ket{\widetilde{\psi}}_1\equiv|a_{00}|\ket{00}+|a_{10}|\ket{01}+|a_{01}|\ket{10}-|a_{11}|\ket{11}$.
Then, we recover the phase factors $e^{i\phi_{ij}}$ by converting
$\ket{\widetilde{\psi}}_1$ to $\ket{\psi}_f\equiv a_{00}\ket{00}+a_{10}\ket{01}+a_{01}\ket{10}-a_{11}\ket{11}$.
Here we will only use the holonomic single- and two-qubit gates achieved in the main text for these two steps.

The procedures for population recovery are listed as follows [see Fig.~\ref{recovery}(b)]:

(i)~We perform a holonomic two-qubit gate,
\[
U_2(\theta_1)=I_{00}\oplus\left(
\begin{array}{cc}
\cos\theta_1 & -\sin\theta_1 \\
-\sin\theta_1 & -\cos\theta_1 \\
\end{array}
\right)\oplus-I_{11},
\]
where $\theta_1=\frac{5}{4}\pi$. After this, $\ket{\psi}_1$ is transformed to
\begin{equation}
  \frac{a_{01}+a_{10}}{a}\ket{00}+\frac{\sqrt{2}a_{11}}{a}\ket{10},
\end{equation}
where $a=\sqrt{|a_{01}+a_{10}|^2+2|a_{11}|^2}$ is a normalization factor and we write
$a_{01}+a_{10}=|a_{01}+a_{10}|e^{i\phi_{01+10}}$.

(ii)~We perform a single-qubit gate,
\begin{align}
U_1=&\left(
\begin{array}{cc}
e^{\frac{i}{2}(\phi_{11}-\phi_{01+10})} & 0 \\
0 & e^{-\frac{i}{2}(\phi_{11}-\phi_{01+10})} \\
\end{array}
\right) \nonumber\\
&\times\left(
\begin{array}{cc}
\cos\theta & \sin\theta e^{i(-\phi_{11}+\phi_{01+10})} \\
\sin\theta e^{i(\phi_{11}-\phi_{01+10})} & -\cos\theta \\
\end{array}
\right)
\end{align}
on the first target qubit, where $\theta=\arccos(|a_{00}|)+\arctan\left(\frac{\sqrt{2}|a_{11}|}{|a_{01}+a_{10}|}\right)$.
Then, the state is transformed to
\begin{equation}\label{}
  |a_{00}|\ket{00}+\sqrt{|a_{01}|^2+|a_{10}|^2+|a_{11}|^2}\ket{10}.
\end{equation}

(iii)~We perform another two-qubit gate,
\[
U_2(\theta_2)=I_{00}\oplus\left(
\begin{array}{cc}
\cos\theta_2 & -\sin\theta_2 \\
-\sin\theta_2 & -\cos\theta_2 \\
\end{array}
\right)\oplus-I_{11},
\]
where $\theta_2=\pi+\arctan\left(\frac{\sqrt{|a_{01}|^2+|a_{10}|^2}}{|a_{11}|}\right)$.
The state is transformed to
\begin{equation}\label{}
  |a_{00}|\ket{00}+\sqrt{|a_{01}|^2+|a_{10}|^2}\ket{01}+|a_{11}|\ket{10}.
\end{equation}

(iv)~We perform a CNOT gate on the two target qubits, which transforms the state to
\begin{equation}\label{}
  |a_{00}|\ket{00}+\sqrt{|a_{01}|^2+|a_{10}|^2}\ket{01}+|a_{11}|\ket{11}.
\end{equation}

(v)~We perform a two-qubit gate,
\[
U_2(\theta_3)=I_{00}\oplus\left(
\begin{array}{cc}
\cos\theta_3 & -\sin\theta_3 \\
-\sin\theta_3 & -\cos\theta_3 \\
\end{array}
\right)\oplus-I_{11},
\]
where $\theta_3=-\arctan\left(\frac{|a_{01}|}{|a_{10}|}\right)$.
The state is then transformed to
\begin{equation}\label{}
  \ket{\widetilde{\psi}}_1=|a_{00}|\ket{00}+|a_{10}|\ket{01}+|a_{01}|\ket{10}-|a_{11}|\ket{11}.
\end{equation}

The procedures for phase recovery are listed as follows [see Fig.~\ref{recovery}(c)]:

(i)~For convenience, we define
\begin{align}
&m=\frac{\phi_{11}+\phi_{01}}{2}-\phi_0,\nonumber\\
&n=\frac{\phi_{11}-\phi_{01}}{2},  \\
&p=\phi_{01}+\phi_{10}-2\phi_0,\nonumber
\end{align}
where $4\phi_0=\phi_{00}+\phi_{01}+\phi_{10}+\phi_{11}$. We perform single-qubit phase-shift gates,
\[
R_1=\left(
\begin{array}{cc}
e^{-im} & 0 \\
0 & e^{im} \\
\end{array}
\right),
~~R_2=\left(
\begin{array}{cc}
e^{-i(n+\frac{p}{2})} & 0 \\
0 & e^{i(n+\frac{p}{2})} \\
\end{array}
\right)
\]
on target qubits 1 and 2, respectively.
After this, $\ket{\widetilde{\psi}}_1$ is transformed to
\begin{align}\label{}
  &e^{-i(m+n+\frac{p}{2})}|a_{00}|\ket{00}+e^{i(-m+n+\frac{p}{2})}|a_{01}|\ket{01}\nonumber\\
  &+e^{i(m-n-\frac{p}{2})}|a_{10}|\ket{10}+e^{i(m+n+\frac{p}{2})}|a_{11}|\ket{11}.
\end{align}

(ii)~We perform a sequence of CNOT, $R^\prime_2$, and CNOT gates, where
\[
R^\prime_2=\left(
                                                                \begin{array}{cc}
                                                                  e^{-i\frac{p}{2}} & 0 \\
                                                                  0 & e^{i\frac{p}{2}} \\
                                                                \end{array}
                                                              \right)
\]
is a phase-shift gate on the target qubit 2. The state is transformed to
\begin{equation}
\ket{\psi}_f=a_{00}\ket{00}+a_{10}\ket{01}+a_{01}\ket{10}-a_{11}\ket{11},
\end{equation}
up to a global phase $e^{-i\phi_0}$. Then, we recover $\ket{\psi^{(i)}}_f\equiv \ket{\psi}_f\otimes\ket{0}$
from $\ket{\psi}_1\otimes\ket{1}$, by mapping $\ket{1}$ of the auxiliary qubit to $\ket{0}$ via a holonomic
single-qubit rotation.

Note that several procedures are needed when recovering $\ket{\psi^{(i)}}_f$ from $\ket{\psi}_1\otimes\ket{1}$,
but actually these will rarely be used because $\ket{\psi^{(r)}}_f$ in Eq.~(\ref{final}) collapses to the state
$\ket{\psi}_1\otimes\ket{1}$ with very low probability when implementing a measurement on the auxiliary qubit with
$\sigma_z$.

\subsection{Measurement with $\sigma_x$}

The real final state in Eq.~(\ref{final}) can also be rewritten as
\begin{align}\label{}
  \ket{\psi^{(r)}}_f=&\frac{1}{\sqrt{2}}[(\ket{\psi}_0+i\sin\delta\ket{\psi}_1)\otimes\ket{+}\nonumber \\
  &+(\ket{\psi}_0-i\sin\delta\ket{\psi}_1)\otimes\ket{-}],
\end{align}
where $\ket{+}$ and $\ket{-}$ are two eigenstates of $\sigma_x$, corresponding to the eigenvalues $+1$ and $-1$,
respectively. By implementing a measurement on the auxiliary qubit with $\sigma_x$, the real final state collapses to
$\ket{\psi^{(r)}}_f=\frac{1}{\sqrt{2}}(\ket{\psi}_0\pm i\sin\delta\ket{\psi}_1)\otimes\ket{\pm}$.
Here each of the states $\ket{\pm}$ of the auxiliary qubit can be reset to $\ket{0}$ by a single-qubit rotation
based on the outcome. When this is performed, the resulting final state is
$\ket{\psi^{(r)}}_f=(\ket{\psi}_0\pm i\sin\delta\ket{\psi}_1)\otimes\ket{0}$, which is close to the ideal final
state in Eq.~(\ref{ideal}) for a small area error $\delta$.

Note that the module of the wave function $\ket{\psi}_0\pm i\sin\delta\ket{\psi}_1$ is
\begin{widetext}
\begin{align}\label{}
  |\ket{\psi}_0\pm i\sin\delta\ket{\psi}_1|= & a_{00}a_{00}^*+i\frac{\sqrt{2}}{2}\sin\delta a_{00}(a_{01}^*
  +a_{10}^*)-i\frac{\sqrt{2}}{2}\sin\delta a_{00}^*(a_{01}+a_{10})+\frac{1}{2}\sin\delta^2(a_{01}
  +a_{10})(a_{01}^*+a_{10}^*)  \nonumber\\
   & +\left(a_{01}\sin^2\frac{\delta}{2}+a_{10}\cos^2\frac{\delta}{2}\right)
   \left(a_{01}^*\sin^2\frac{\delta}{2}+a_{10}^*\cos^2\frac{\delta}{2}\right)
   +i\frac{\sqrt{2}}{2}\sin\delta a_{11}\left(a_{01}^*\sin\frac{\delta}{2}+a_{10}^*\cos\frac{\delta}{2}\right) \nonumber \\
   &-i\frac{\sqrt{2}}{2}\sin\delta a_{11}^*\left(a_{01}\sin\frac{\delta}{2}+a_{10}\cos\frac{\delta}{2}\right)
   +\frac{1}{2}\sin^2\delta a_{11}a_{11}^* \nonumber \\
   & +\left(a_{10}\sin^2\frac{\delta}{2}+a_{01}\cos^2\frac{\delta}{2}\right)
   \left(a_{10}^*\sin^2\frac{\delta}{2}+a_{01}^*\cos^2\frac{\delta}{2}\right)
   +i\frac{\sqrt{2}}{2}\sin\delta a_{11}\left(a_{10}^*\sin\frac{\delta}{2}+a_{01}^*\cos\frac{\delta}{2}\right) \nonumber \\
   &-i\frac{\sqrt{2}}{2}\sin\delta a_{11}^*\left(a_{10}\sin\frac{\delta}{2}+a_{01}\cos\frac{\delta}{2}\right)
   +\frac{1}{2}\sin^2\delta a_{11}a_{11}^*+a_{11}a_{11}^*\cos^2\delta.
\end{align}
\end{widetext}
After the measurement on the auxiliary qubit with $\sigma_x$, the fidelity of the two-qubit gate can be written as
\begin{align}\label{}
  F=&\frac{|\bra{\psi^{(i)}}(\ket{\psi}_0\pm i\sin\delta\ket{\psi}_1)\otimes\ket{0}|}
  {|\ket{\psi}_0\pm i\sin\delta\ket{\psi}_1|} \nonumber \\
  =&\frac{1}{|\ket{\psi}_0\pm i\sin\delta\ket{\psi}_1|}[a_{00}a_{00}^*
  +\sin^2\frac{\delta}{2}(a_{10}^* a_{01}+a_{01}^* a_{10})\nonumber\\
  &+\cos^2\frac{\delta}{2}(a_{10}^* a_{10}+a_{01}^* a_{01})+a_{11}^*a_{11}\cos\delta\nonumber \\
  &+i\frac{1}{\sqrt{2}}\sin\delta(a_{00}^*a_{01}+a_{00}^*a_{10}+a_{10}^*a_{11}+a_{01}^*a_{11})].
\end{align}
As an estimation, we consider the case with real $a_{ij}$'s and then have
\begin{align}\label{}
  F=&\frac{1-\frac{\delta^4}{24}|a_{01}a_{10}|}{1-\frac{5}{24}\delta^4(a_{10}^2+a_{01}^2+2a_{10}a_{01})},
\end{align}
where both $\sin\delta\approx\delta$ and $\cos\delta\approx1-\frac{\delta^2}{2}$ are taken.
Using the lower bound condition $a_{01}=a_{10}=\frac{1}{\sqrt{2}}$, we obtain the lower bound of the two-qubit gate
fidelity
\begin{equation}\label{}
  F_h=\frac{1-\frac{\delta^4}{48}}{1-\frac{5}{12}\delta^4}\approx1-\frac{19}{48}\delta^4,
\end{equation}
which is larger than $F_h=1-\frac{\delta^4}{1-\delta^2+\delta^4}\approx 1-\delta^4$ in Eq.~(\ref{hfide1}).
Thus, this measurement improves the fidelity of the two-qubit gate.

\end{document}